\documentclass[preprint,12pt]{elsarticle}

\usepackage{tikz}
\usetikzlibrary{backgrounds}
\usetikzlibrary{shapes.geometric, arrows}
\usepackage{amssymb}
\usepackage{amsmath}
\usepackage[normalem]{ulem}
\biboptions{sort&compress}
\usepackage{enumitem}
\usepackage{siunitx}
\usepackage[margin=1in]{geometry}
\usepackage[export]{adjustbox}
\usepackage{lineno,hyperref}
\hypersetup{colorlinks = true,linkcolor = blue,anchorcolor =red,citecolor = blue,filecolor = red,urlcolor = red,
            pdfauthor=author}
\modulolinenumbers[1]
\usepackage{amsmath,mathtools,mathrsfs,amsfonts}

\usepackage{comment}
\usepackage{cleveref}
\usepackage{graphicx,bm,algpseudocode}
\usepackage{subcaption,caption}
\usepackage{multirow}
\usepackage{miller}
\journal{Scripta Materialia}

\usepackage{xcolor}
\definecolor{C0}{HTML}{1F77B4}
\definecolor{C1}{HTML}{FF7F0E}
\definecolor{C2}{HTML}{2ca02c}
\definecolor{C3}{HTML}{d62728}
\definecolor{C4}{HTML}{9467bd}
\definecolor{C5}{HTML}{8c564b}
\def\usetodonotes{} 
\ifdefined\usetodonotes
\usepackage[colorinlistoftodos]{todonotes}
\usepackage{xparse}
\usepackage{fontawesome5}
\NewDocumentCommand{\brandon}{O{} m}{\todo[color=C0!10,linecolor=C0,bordercolor=C0,#1]{BR: #2}}

\usepackage{xpatch,environ,ragged2e,setspace}
\RenewEnviron{abstract}{%
  \xdef\theabstracttext{%
    \unexpanded{%
      \def\baselinestretch{2}\noindent\unskip\textbf{Abstract}\par\medskip
      \noindent\unskip\ignorespaces}%
    \unexpanded\expandafter{\BODY}%
  }%
}
\def\theabstracttext{}
\xpatchcmd{\pprintMaketitle}
  {\ifvoid\absbox}
  {\ifx\theabstracttext\empty\else\printtheabstracttext\fi\ifvoid\absbox}
  {}{}
\newcommand{\printtheabstracttext}{{%
  \begin{trivlist}
  \normalfont\normalsize
  \item\relax
  \doublespacing\theabstracttext
  \end{trivlist}
}}
\fi

\usepackage[commandnameprefix=always,commentmarkup=todo,authormarkup=none]{changes}
\definechangesauthor[color=C0]{brandon}
\definechangesauthor[color=C1]{nikhil}
\definechangesauthor[color=C2]{ian}
\definechangesauthor[color=C3]{himanshu}

\makeatletter
\def\ps@pprintTitle{%
  \let\@oddhead\@empty
  \let\@evenhead\@empty
  \def\@oddfoot{\reset@font\hfil\thepage\hfil}
  \let\@evenfoot\@oddfoot
}
\makeatother

\journal{Acta Materialia}

\begin{document}

\begin{frontmatter}

\title{Energetics of the nucleation and glide of disconnection modes in symmetric tilt grain boundaries}

\author[mymainaddress]{Himanshu Joshi\corref{mycorrespondingauthor}}

\author[mytertiaryaddress]{Ian Chesser}
\author[isu]{Brandon Runnels}
\author[mymainaddress]{Nikhil Chandra Admal}
\cortext[mycorrespondingauthor]{Corresponding author}

\address[mymainaddress]{Department of Mechanical Science and Engineering, University of Illinois at Urbana-Champaign, Urbana, IL USA}
\address[mytertiaryaddress]{X Computational Physics Division, Los Alamos National Laboratory, Los Alamos, New Mexico, NM USA}
\address[isu]{Department of Aerospace Engineering, Iowa State University, Ames, IA USA}
\begin{abstract}

GBs evolve by the nucleation and glide of disconnections, which are dislocations with a step character. 
In this work, motivated by recent success in predicting GB properties such as the shear coupling factor and mobility from the intrinsic properties of disconnections,
 
we develop a systematic method to calculate the energy barriers for the nucleation and glide of individual disconnection modes under arbitrary driving forces. This method combines tools from bicrystallography to enumerate disconnection modes and the NEB method to calculate their energetics, yielding minimum energy paths and atomistic mechanisms for the nucleation and glide of each disconnection mode. We apply the method to accurately predict shear coupling factors of \hkl[001] symmetric tilt grain boundaries in Cu. Particular attention is paid to the boundaries where the classical disconnection nucleation model produces incorrect nucleation barriers. We demonstrate that the method can accurately compute the energy barriers and predict the shear coupling factors for low temperature regime.

\end{abstract}
\begin{keyword}
Grain boundary plasticity, Disconnections, Bicrystallography, Energy barriers, Nudged elastic band.
\end{keyword}

\end{frontmatter}

\section{Introduction}
\label{sec:Introduction}
Nanocrystalline materials have many exceptional mechanical and chemical properties compared to conventional materials \cite{lehockey1998improving}. The use of such materials as catalysts \cite{trudeau1996nanocrystalline}, Li-ion batteries \cite{wang2000Liion} and under extreme thermomechanical conditions makes a thorough understanding of mechanisms of plasticity in nano-crystalline materials imperative.  While dislocations-based mechanisms explain the bulk plasticity in such materials, recent experimental and molecular dynamics (MD) results show that disconnections describe the elementary mechanisms for grain boundary (GB) migration \cite{zhu2019situ,zhu2022atomistic,deng2017size}.

A disconnection $(\bm b, h)$ is a GB line defect with a Burgers vector $\bm{b}$ and a step height $h$. The nucleation of a disconnection loop in a GB and its subsequent expansion results in the translation of the GB by the step height $h$, and the region swept by the GB undergoes a plastic shear of magnitude equal to the shear coupling factor $b/h$. Every GB has a countably infinite number of disconnections \cite{admalSNF2022}, called disconnection modes. \citet{khater2012disconnection} proposed the \emph{classical disconnection nucleation model} wherein a disconnection is modeled as discrete dislocation with a step character, and the nucleation rate of a disconnection dipole is determined by the energy barrier computed using the classical dislocation theory adapted to disconnections. Consequently, the nucleation of disconnections with the lowest energy barrier is the most favorable and therefore determines the shear coupling factor. \citet{han2018grain} showed that for most symmetric tilt grain boundaries (STGBs), the shear coupling factors predicted by the classical disconnection nucleation model are in reasonable agreement with those observed in MD simulations of \citet{homer2013}. However, some exceptions were noted especially at the transition region between distinct coupling modes. Furthermore, in an earlier work \cite{joshi_atgb}, we adapted the classical disconnection nucleation model to asymmetric tilt grain boundaries (ATGBs) and observed that the model does not perform as well as it does for STGBs. 

The above-stated limitations may be attributed to the continuum dislocation theory's inaccurate prediction of the nucleation barrier. This can be a result of the classical model's inability to account for interface structure and atomic shuffles during GB motion. This motivates us to employ atomistics to systematically map the nucleation barriers of individual disconnection modes. Evidently, the nucleation barrier depends on the nature of the driving force (shear/chemical potential) \cite{chen2019grain,han2018grain}. In recent work, \citet{rajabzadeh2013elementary} and \citet{combe2016disconnections} used MD to study the atomistic mechanism of disconnection nucleation and glide during shear-driven GB motion and subsequently used the nudged elastic band (NEB) method to calculate the associated energy barrier. However, their calculation was limited to the disconnection mode observed in the MD simulation, and therefore, a single such calculation cannot be used to predict the shear coupling factor for different magnitudes of driving forces attributed to shear and/or chemical potential. This paper aims to develop a framework for calculating the energetics of the nucleation and glide of individual disconnection modes and subsequently predict the shear coupling of a GB subject to \emph{any} driving force.

The paper is organized as follows:
\Cref{sec:Methods} presents tools to explore STGBs disconnection modes and calculate the nucleation and glide energy barriers using NEB. 
In \Cref{sec:results}, we present and compare the results of NEB calculations. 
We summarize and conclude in \Cref{sec:Conclusions}.

\section{Methodology}
\label{sec:Methods}
In this section, we describe the steps to construct disconnections in STGBs and to calculate the energy barriers associated with their nucleation and glide using the NEB method of LAMMPS \cite{lmp}. In addition, to validate the shear coupling factors predicted from the calculated energy barriers, we also conduct molecular dynamics (MD) simulations of STGB migration using LAMMPS. The methodology is implemented for Copper using the EAM potential \citep{mishin2001eam}.

\subsection{STGB energy barrier calculation}
The methodology adopted in this study can be broken down into 5 steps, described below: 

\paragraph{\textbf{Step 1:} \uline{Construction of STGB microstate}}
\label{sec:stgb_construction}
We follow the standard methodology to create GB microstructure. Each STGB is constructed by rotating two lattices by $\theta/2$ and $-\theta/2$, where $\theta$ is the misorientation angle. 
We refer to the resulting STGB as a \emph{perfect GB}, which is typically \emph{not} in the lowest energy state. 
To arrive at the ground state, we use the $\gamma$-surface method, wherein the two grains are displaced relative to each other in the GB plane and energy is minimized. 
The relative displacement that results in the lowest post-minimized energy is recorded.
\begin{figure}[h!]
    \centering
    \begin{subfigure}[b]{0.48\textwidth}
        \frame{\includegraphics[width=\linewidth,trim={16cm 18cm 16cm 18cm},clip]{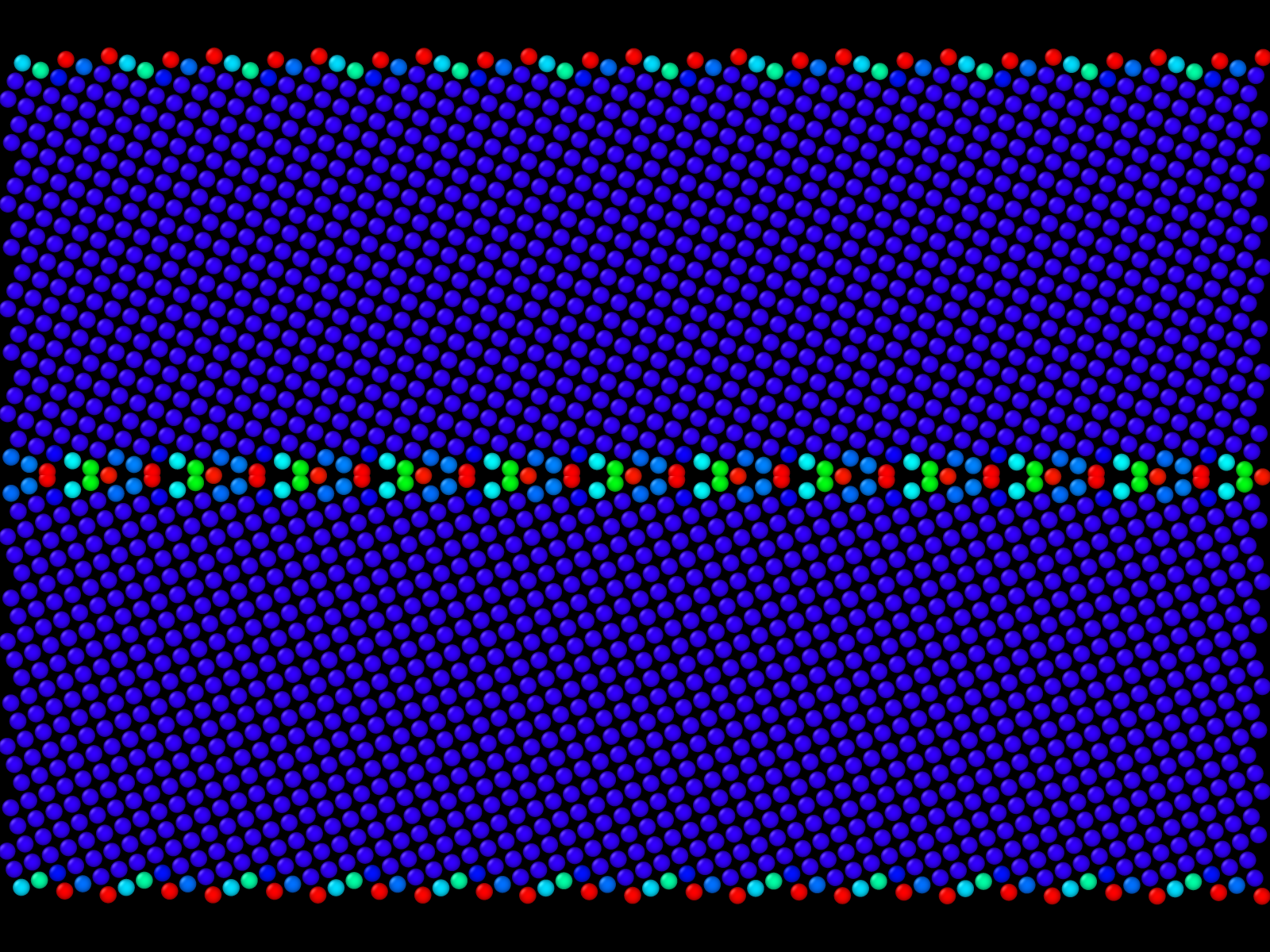}}
        \caption{GB from bicrystallography}
        \label{fig:sigma29_ascut} 
    \end{subfigure}
     \begin{subfigure}[b]{0.48\textwidth}
        \frame{\includegraphics[width=\linewidth,trim={16cm 17.5cm 16cm 18.5cm},clip]{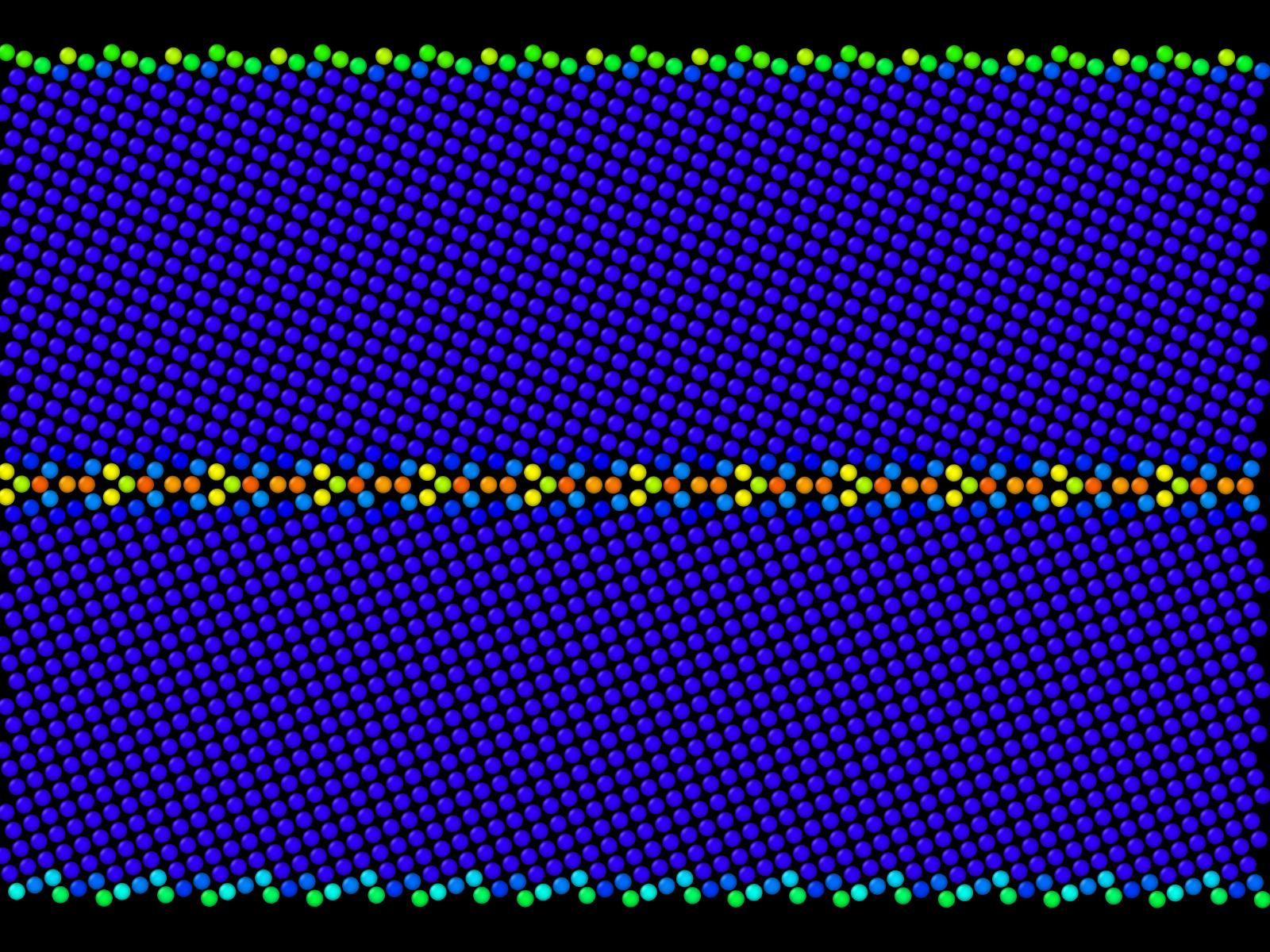}}
        \caption{GB at lowest energy state}
        \label{fig:sigma29_gammasurface} 
    \end{subfigure}
    \caption{GB microstate generated for $\Sigma29\hkl[0 0 1]\hkl(5 2 0)$ using bicrytallography and application of gamma surface method. Atoms are colored according to the centrosymmetry parameter, visualized using OVITO\cite{ovito}. 
    }
    \label{fig:stgb_microstates}
\end{figure}

\paragraph{\textbf{Step 2:} \uline{Enumeration of disconnection modes using SNF bicrystallography}}
\label{sec:snf}
We use SNF bicrystallography to enumerate the disconnection modes $(\bm b,h)$ of an STGB, where $\bm b$ and $h$ denote the disconnection's Burgers vector and step height, respectively. SNF bicrystallography is a powerful framework based on integer matrix algebra to automate the generation of rational GBs, and enumerate disconnection modes in GBs. The SNF framework results in dimension-independent algorithms that apply to any crystal system. \footnote{SNF bicrystallography is implemented as an opensource C++ library \emph{open Interface Lab} (\texttt{oiLAB}), accessible at \url{https://github.com/oilab-project/oILAB.git}.} 
For further details on SNF bicrystallography, we refer the reader to \cite{admalSNF2022}. 

\begin{figure}[h!]
    \centering
    \includegraphics[width=\linewidth,height=3.5cm]{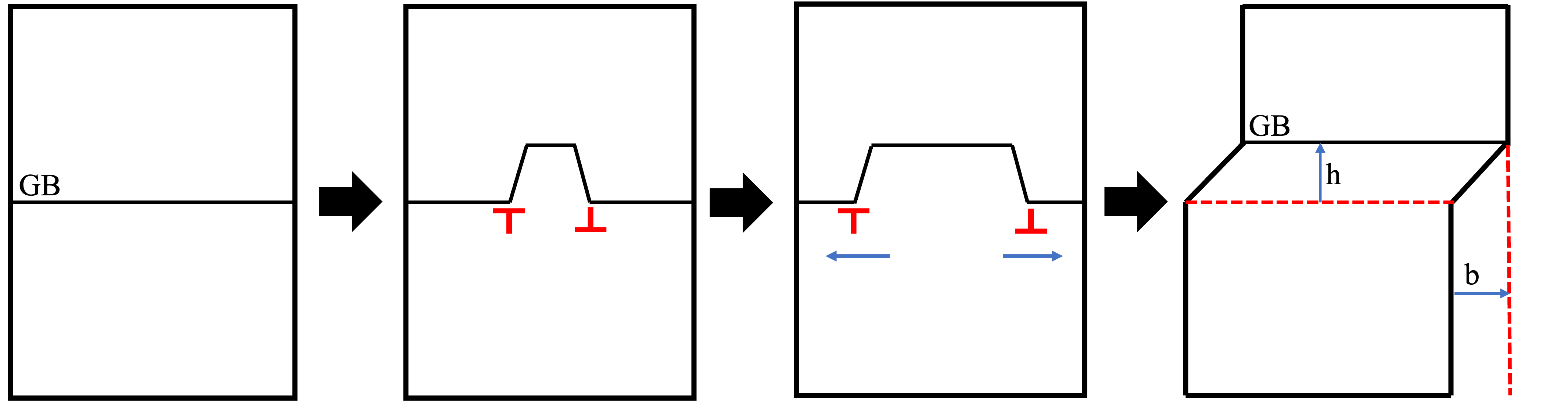}
    \caption{A schematic of migration of GB with disconnection nucleation and glide as the underlying mechanism. We start with a flat GB, due to applied stress a disconnection is nucleated with a step height (h) and burgers vector (b). The disconnection glides across the GB leading to GB migration.}
    \label{fig:gb_migration_cartoon}
\end{figure}

\paragraph{\textbf{Step 3:} \uline{Construction of atomistic GB migration images using disconnections}}
This step concerns the construction of atomistic images of GB migration in a bicrystal with disconnections as the underlying mechanism. We assume that GB migrates due to nucleation of a disconnection (represented by a dislocation dipole nucleating in the GB) and its subsequent glide across the GB. Figure \ref{fig:gb_migration_cartoon} shows a simplified representation of GB migration using disconnections.

\begin{figure}
    \centering
    \begin{subfigure}[b]{0.48\textwidth}
        \centering
        \includegraphics[width=0.75\linewidth]{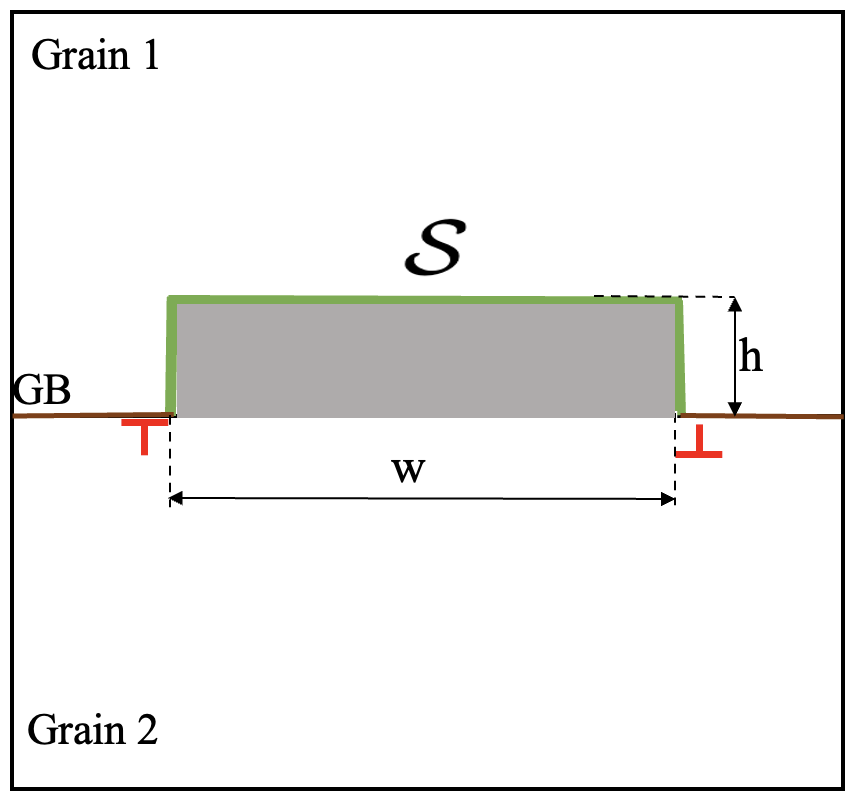}
        \caption{}
        \label{fig:dc_1} 
    \end{subfigure}
    \quad
     \begin{subfigure}[b]{0.48\textwidth}
        \centering
        \includegraphics[width=\linewidth]{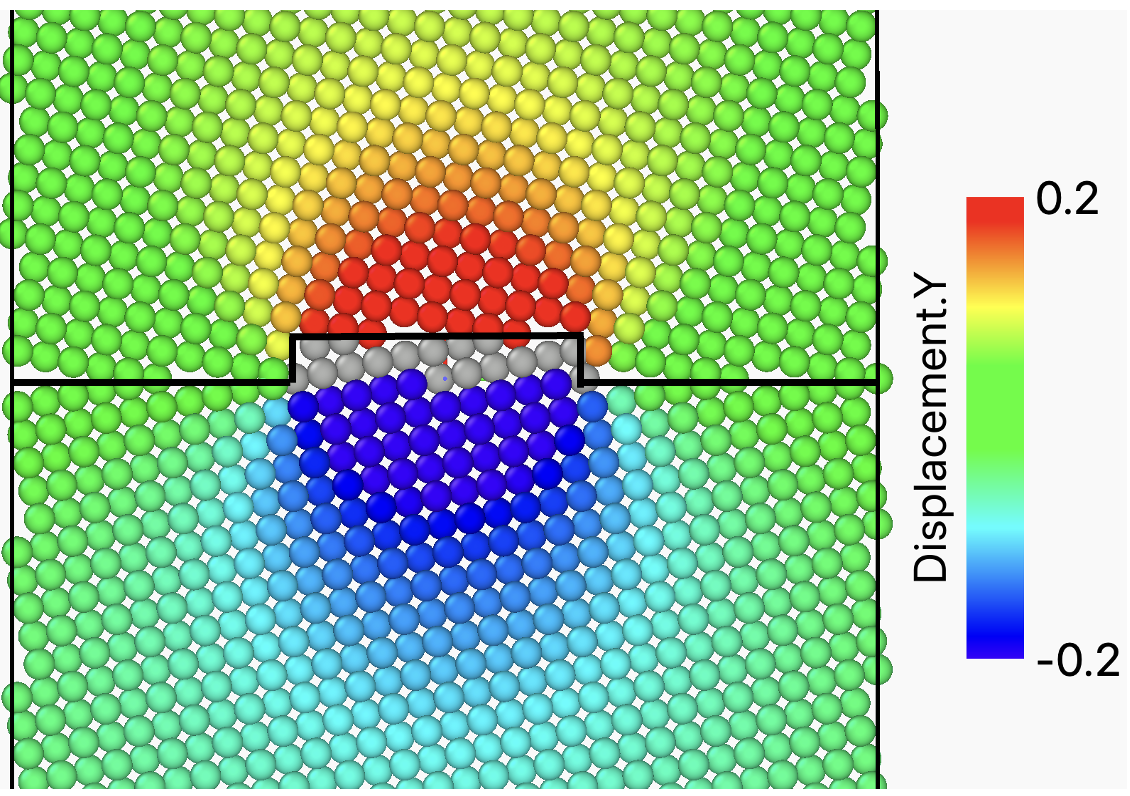}
        \caption{}
        \label{fig:dc_2} 
    \end{subfigure}
    \caption{An atomistic construction of a disconnection dipole. (a) shows a schematic of a disconnection dipole with a slip surface $\mathcal S$ highlighted in green. (b) shows the resulting atomistic system obtained by displacing atoms according to  \Cref{eqn:plastic_displacement_field} for $\Sigma 73[001](11\ 5\ 0)$. The atoms which transform from grain 1 orientation to grain 2 orientation are colored grey. The atoms present outside the transformed region are colored according to the displacement of atoms along the GB.}
    \label{fig:disconnection}
\end{figure}

For each of the images we construct, we begin with the dichromatic pattern formed by two interpenetrating lattices, rotated by $\theta/2$ and $-\theta/2$. To construct a disconnection dipole $(\bm b,h)$ of a prescribed width $w$, as shown in  \Cref{fig:dc_1}, all atoms in the dichromatic pattern are displaced parallel to the GB according to the plastic displacement field
\begin{equation}
    u_(\bm{x};\mathcal S) = -\frac{\bm{b}\Omega(\bm{x};\mathcal S)}{4\pi},
    \label{eqn:plastic_displacement_field}
\end{equation}
where the surface $\mathcal S$, shown in green in \Cref{fig:dc_1}, denotes the slip surface, and $\Omega(\bm{x};\mathcal S)$ is the solid angle subtended at a point $\bm x$ by $\mathcal S$.
Next, the stepped GB is formed by deleting atoms of lattices on either side of $\mathcal S$. \Cref{fig:dc_2} shows the displacement of atoms and highlights the a jump discontinuity of $|\bm b|$ across $\mathcal S$ for $\Sigma 73[001](11\ 5\ 0)$. 

\begin{figure}[h!]
    \centering
     \begin{subfigure}[b]{0.48\textwidth}
        \frame{\includegraphics[width=\textwidth]{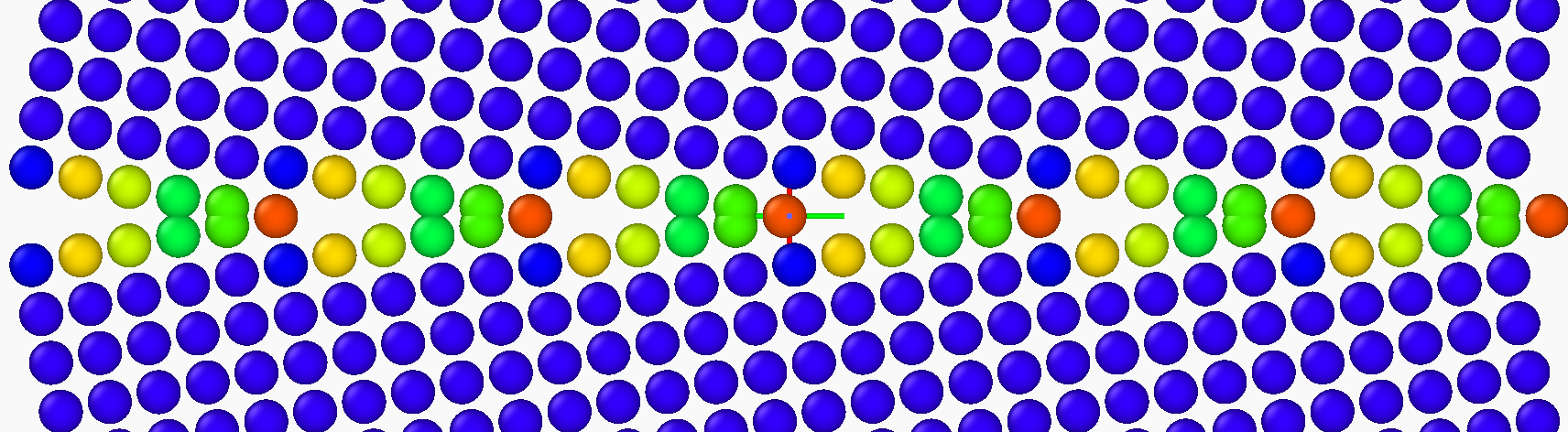}}
        \caption{}
        \label{fig:images-1} 
    \end{subfigure}
    \quad
     \begin{subfigure}[b]{0.48\textwidth}
        \frame{\includegraphics[width=\textwidth]{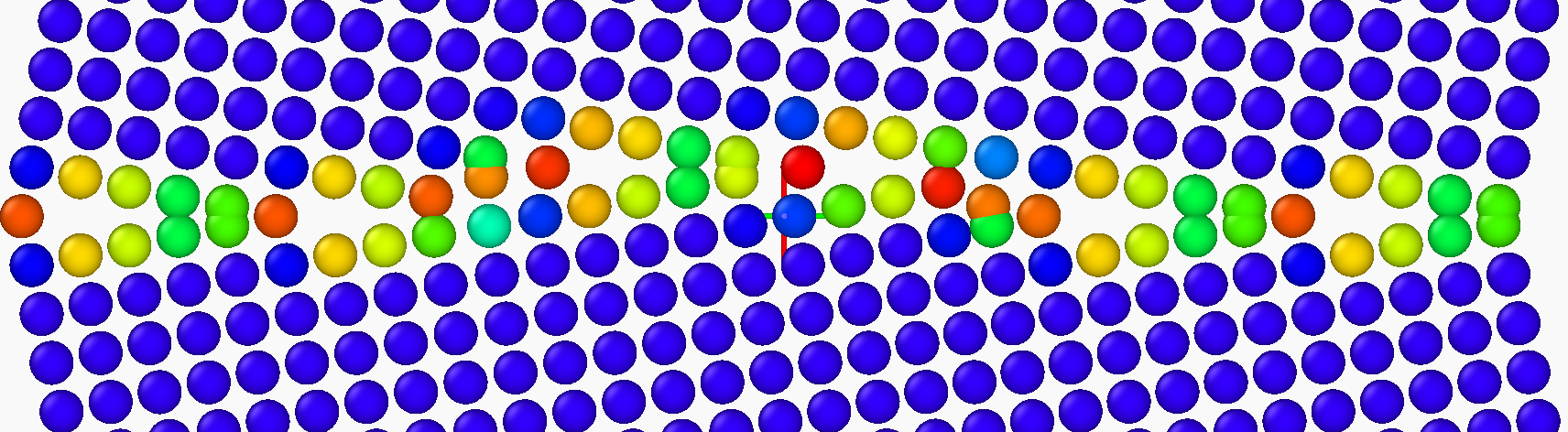}}
        \caption{}
        \label{fig:images-2} 
    \end{subfigure}
    \hfill
     \begin{subfigure}[b]{0.48\textwidth}
        \frame{\includegraphics[width=\textwidth]{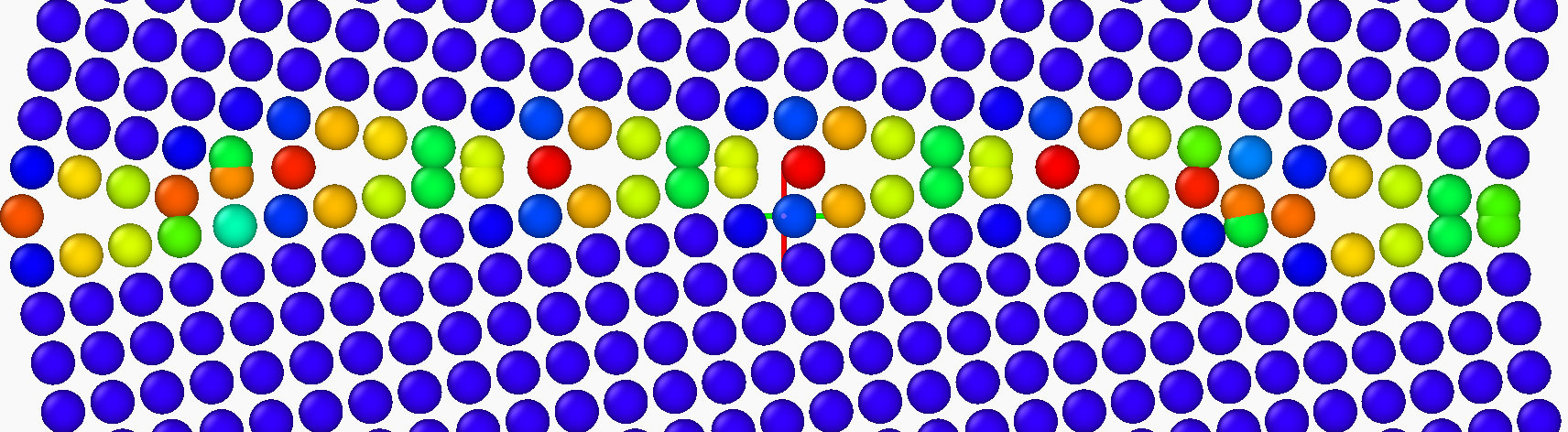}}
        \caption{}
        \label{fig:images-3} 
    \end{subfigure}
    \quad
     \begin{subfigure}[b]{0.48\textwidth}
        \frame{\includegraphics[width=\textwidth]{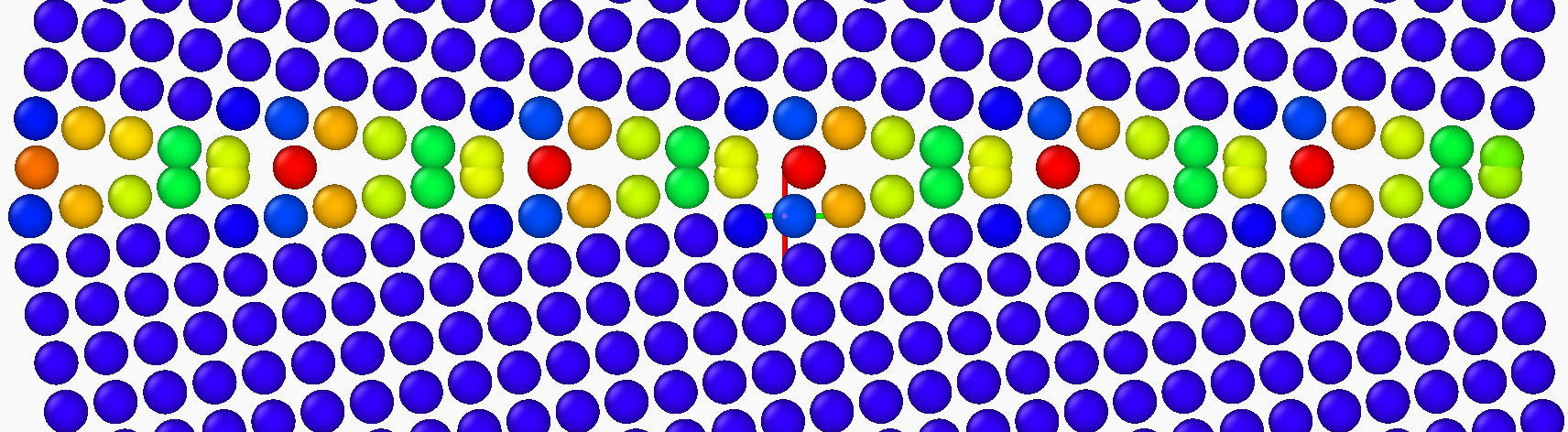}}
        \caption{}
        \label{fig:images-4} 
    \end{subfigure}
    \caption{GB structures in a $\Sigma 13[001](5\ 1\ 0)$ STGB, constructed using bicrytallography. (a) shows the initial GB configuration, (b) shows a disconnection nucleated on the GB, (c) shows the GB configuration as the disconnection glides on the GB and (d) shows final GB configuration. Particles are colored according to the centrosymmetry parameter \cite{ovito}.}
    \label{fig:sigma13_nonminimized_images}
\end{figure}

We repeat the method described above for increasing step widths (each subsequent image having a step width increased by 1 CSL period up to the GB length along the direction perpendicular to the tilt axis) resulting in a sequence of configurations, as shown in Figure \ref{fig:sigma13_nonminimized_images} corresponding to the disconnection mode $(-0.7,1.77)$ for $\Sigma 13[001](5\ 1\ 0)$ GB. The flat GB in \Cref{fig:images-1} eventually translates by the step height $h$ to form the flat GB in \Cref{fig:images-4}.
Note that since glide disconnection is volume-preserving, all configurations in \Cref{fig:sigma13_nonminimized_images} have equal number of atoms.

\begin{figure}[h!]
    \centering
     \begin{subfigure}[b]{0.48\textwidth}
        \frame{\includegraphics[width=\textwidth]{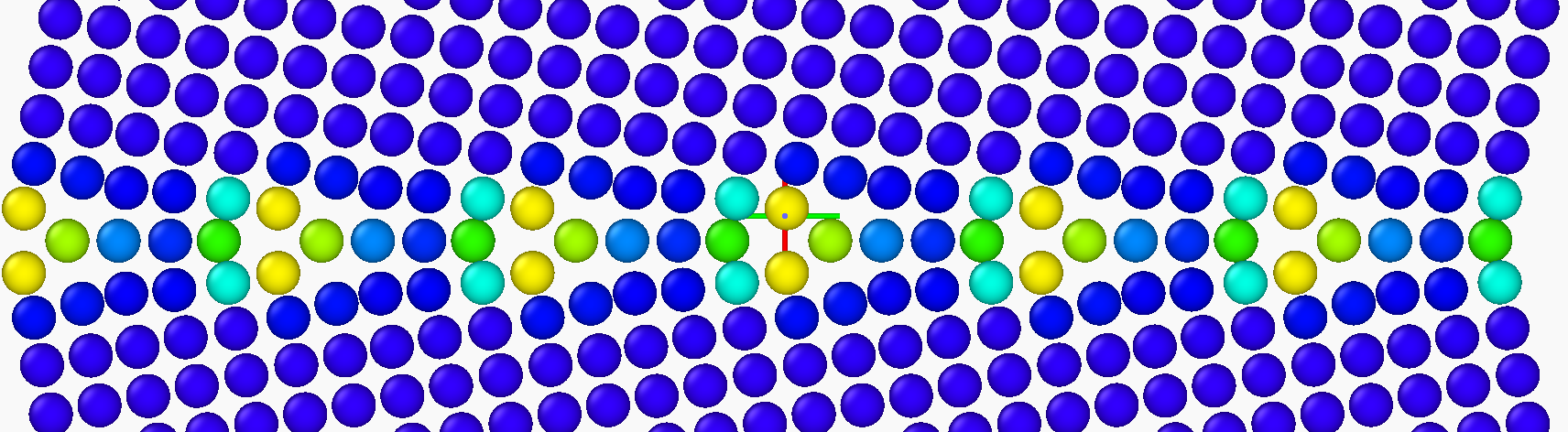}}
        \caption{}
        \label{fig:images-5} 
    \end{subfigure}
    \quad
     \begin{subfigure}[b]{0.48\textwidth}
        \frame{\includegraphics[width=\textwidth]{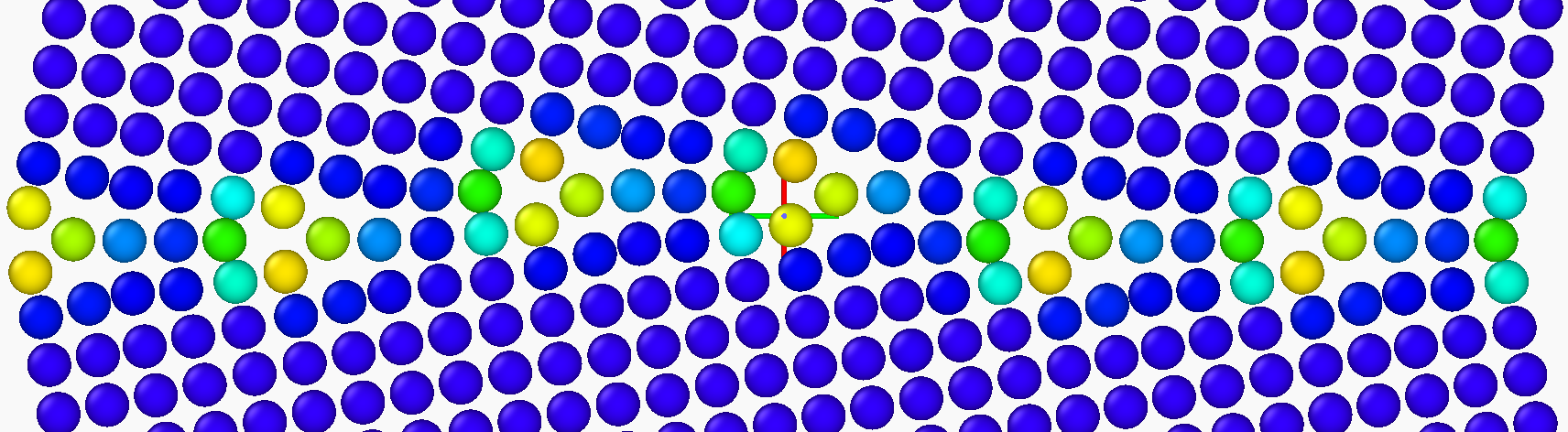}}
        \caption{}
        \label{fig:images-6} 
    \end{subfigure}
    \hfill
     \begin{subfigure}[b]{0.48\textwidth}
        \frame{\includegraphics[width=\textwidth]{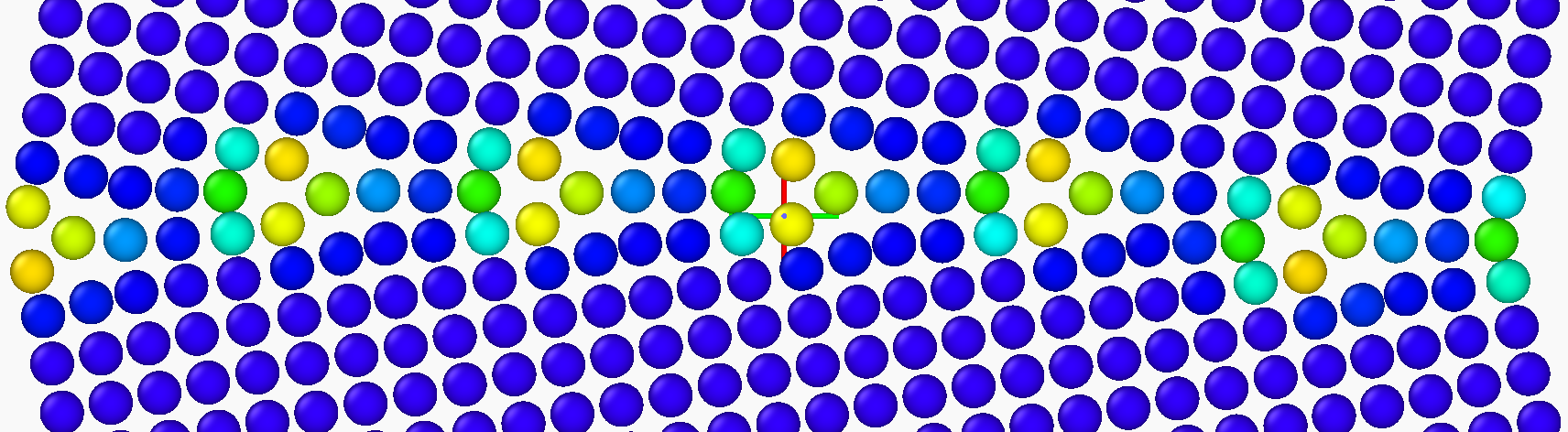}}
        \caption{}
        \label{fig:images-7} 
    \end{subfigure}
    \quad
     \begin{subfigure}[b]{0.48\textwidth}
        \frame{\includegraphics[width=\textwidth]{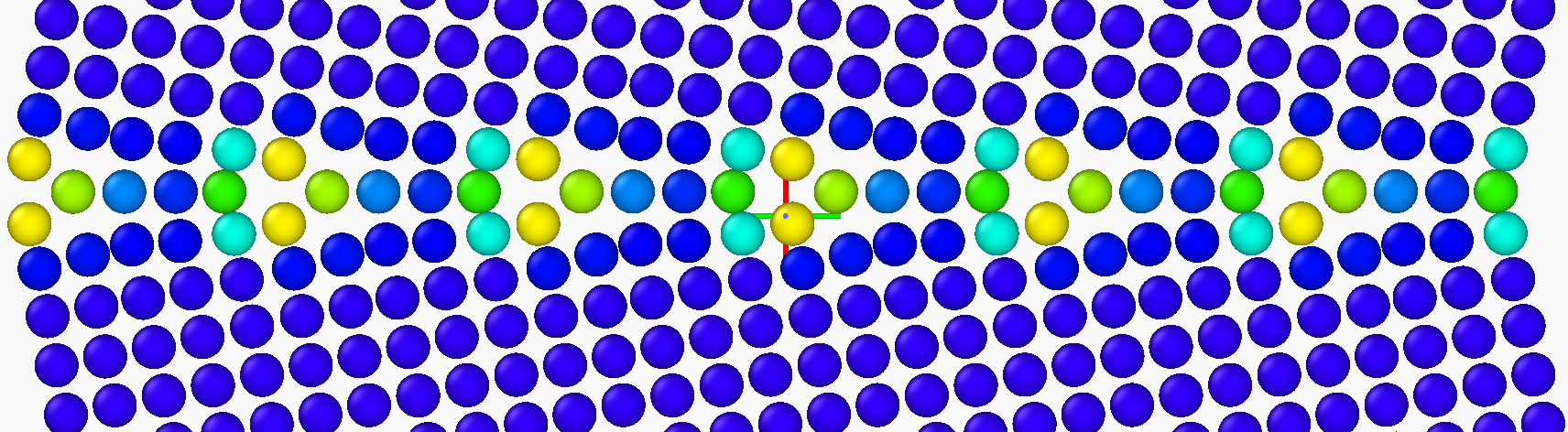}}
        \caption{}
        \label{fig:images-8} 
    \end{subfigure}
    \caption{Relaxed GB structures in a $\Sigma 13[001](5\ 1\ 0)$  STGB, obtained in Step 3. (a) shows the initial GB configuration, (b) shows a disconnection nucleated on the GB, (c) shows the GB configuration as the disconnection glides on the GB and (d) shows final GB configuration. The atoms are colored by centrosymmetry parameter to highlight the GB \cite{ovito}.} 
    \label{fig:images}
\end{figure}

The GB microstates in images thus created, while not in the lowest energy state, are identical as a disconnection preserves the GB structure. To transform to the lowest energy microstate, we translate one lattice relative to the other along the GB plane recorded in step 1. For STGBs, we have multiple displacements that result in the lowest energy microstate, but we select the one with the displacement predominantly along the tilt axis. Subsequently, since the plastic displacement field in \Cref{eqn:plastic_displacement_field} does not include elastic relaxation, the system is relaxed resulting in disconnections with the ground state GB structure, as shown in \Cref{fig:images}.

We observed that if the step width is too narrow, the stepped GB becomes unstable and snaps back to its flat state. Additionally, in disconnection modes with large Burgers vectors (those greater than twice the smallest Burgers vector associated with the GB), the step becomes unstable and leads to irregular GB structures after minimization. In these cases, we reapply the gamma surface method on the stepped GB to determine if it results in a new relative displacement. The minimized images have a significant impact on the accurate determination of atomic shuffles in Step 4. Therefore, Step 3 necessitates careful inspection to ensure the disconnection steps in intermediate images remain stable.

\begin{figure}[h!]
    \centering
     \begin{subfigure}[b]{0.3\textwidth}
        \frame{\includegraphics[width=\textwidth]{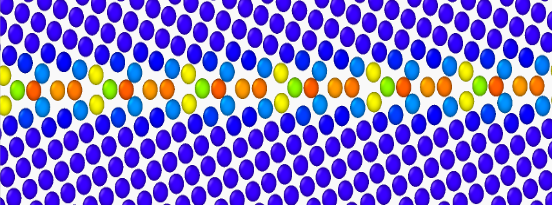}}
        \caption{}
        \label{fig:mapping_1-1} 
    \end{subfigure}
    \quad
    \begin{subfigure}[b]{0.3\textwidth}
        \frame{\includegraphics[width=\textwidth]{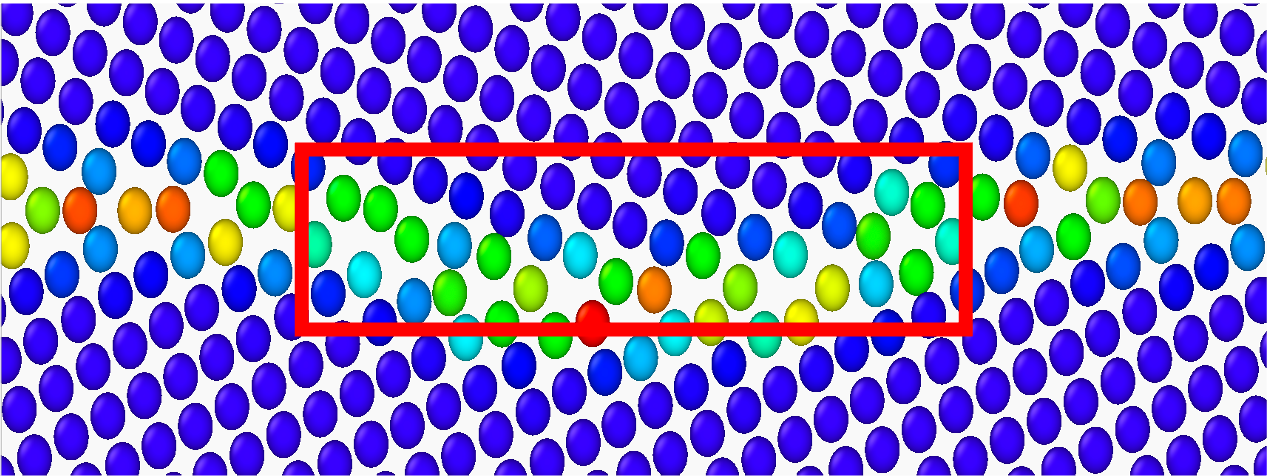}}
        \caption{}
        \label{fig:mapping_1-2} 
    \end{subfigure}
    \quad 
    \begin{subfigure}[b]{0.3\textwidth}
    \frame{\includegraphics[width=\linewidth,trim={0cm 0cm 0cm 0.5cm},clip]{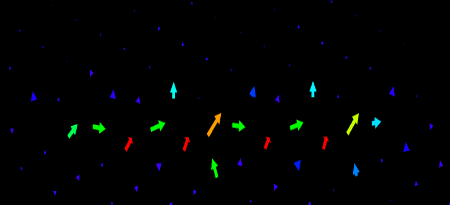}}
        \caption{}
        \label{fig:mapping_1-3} 
    \end{subfigure}
    \hfill
     \begin{subfigure}[b]{0.3\textwidth}
        \frame{\includegraphics[width=\textwidth]{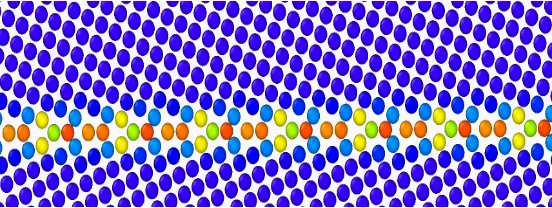}}
        \caption{}
        \label{fig:mapping_2-1} 
    \end{subfigure}
    \quad
    \begin{subfigure}[b]{0.3\textwidth}
        \frame{\includegraphics[width=\textwidth]{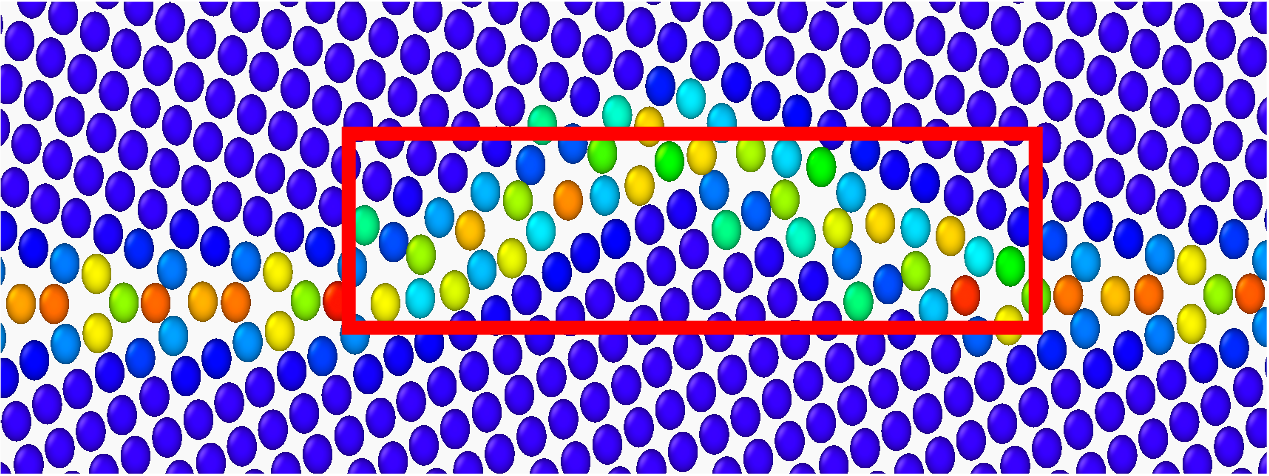}}
        \caption{}
        \label{fig:mapping_2-2} 
    \end{subfigure}
    \quad 
    \begin{subfigure}[b]{0.3\textwidth}
    \frame{\includegraphics[width=\linewidth,trim={0cm 0.25cm 0cm 0.25cm},clip]{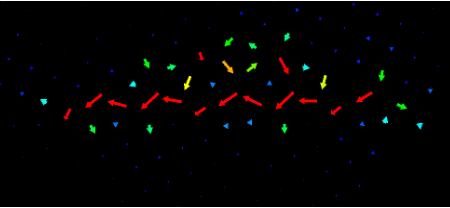}}
        \caption{}
        \label{fig:mapping_2-3} 
    \end{subfigure}
    \caption{Mapping of atomic trajectories using min-shuffle algorithm for different disconnection modes in $\Sigma 29[001]\hkl(5 2 0)$ STGB.(a) shows the initial flat GB for mode $(b,h) = (0,67,-4.03)$, (b) shows a disconnection inserted into the GB for $(b,h) = (0,67,-4.03)$ ,(c) shows the mapping of atoms in the region highlighted in red in (b) for $(b,h) = (0,67,-4.03)$; (d) shows the initial flat GB, (e) shows a disconnection inserted into the GB $(b,h) = (0,67,5.71)$,(f) shows the mapping of atoms in the region highlighted in red in (e) for $(b,h) = (0,67,5.71)$. Atoms are colored according to the centrosymmetry parameter. The arrows depict atomic displacements, with colors representing displacement magnitude.} 
    \label{fig:mapping}
\end{figure}

\paragraph{\textbf{Step 4:} \uline{Mapping the atomic shuffles using optimal transport}}
In this step, we compute the permutation/shuffle maps between the atoms in the flat GB and each of the stepped GB configurations from Step 3. Such maps are well-defined since all the configurations have an identical number of atoms and serve as \emph{intermediate} maps for NEB calculation in Step 5. 
Therefore, Step 4 connects the images generated in Step 3, resulting in a trajectory in the configuration space that will be inputted into the NEB in Step 5. Shuffle maps are calculated using the optimal transport method developed by \citet{chesser2022taxonomy,chesser2021optimal}. The chosen shuffle map, the \emph{min-shuffle} map, minimizes net shuffle distance in the dichromatic pattern. By increasing the regularization parameter described in \cite{chesser2021optimal}, it is possible to select for mappings with larger net displacements such as those observed at high temperatures in specific GBs \cite{chesser2022taxonomy}. In this work, we restrict our attention to the min-shuffle map typically observed in MD simulations of GB migration at low and intermediate temperatures \cite{chesser2022taxonomy}. We use \citet{flamary2021pot} in our implementation of min shuffle calculations. \Cref{fig:mapping} shows the mapping computed using this method for $\Sigma 29\hkl[0 0 1]\hkl(5 2 0)$ STGB. Images \Cref{fig:mapping_1-1} and \Cref{fig:mapping_2-1} show the initial flat GBs, \Cref{fig:mapping_1-2} and \Cref{fig:mapping_2-2} shows the atom positions with a disconnection step inserted into the GB for disconnection modes $(b,h) = (0,67,-4.03)$ and $(b,h) = (0,67,5.71)$ ,respectively.  \Cref{fig:mapping_1-3} and \Cref{fig:mapping_2-3} show the displacement of atoms in the area highlighted in red box in \Cref{fig:mapping_1-2} and \Cref{fig:mapping_2-2},respectively . 

\paragraph{\textbf{Step 5:} \uline{Evaluation of the energy barrier and trajectories using NEB}}

We employ climbing image NEB method \citep{henkelman2000climbing} to calculate the \emph{minimum energy paths (MEPs)}, i.e. plots of energy versus the reaction coordinate (width or the normalized width of the disconnection dipole), of disconnection modes in the absence of external loads. We use $40$ intermediate configurations and mappings from Steps 3 and 4 as inputs to NEB. It is important to note that NEB can be implemented without manually inputting the intermediate images. However, the intermediate images from steps 3 and 4 play a critical role in identifying the lowest energy barrier. The NEB is implemented using the LAMMPS \texttt{neb} module with nudging forces parallel and perpendicular to the configurational path with a unitary spring constant.

Steps 1-5 are implemented in a Python code, available at the GitHub repository \emph{GB\_kinetics} archived at \url{https://github.com/hjoshi9/GB_kinetics.git}.

\subsection{MD simulations}
\label{sec:MD_setup}
\begin{figure}[h!]
    \centering
    \includegraphics[width=0.85\linewidth]{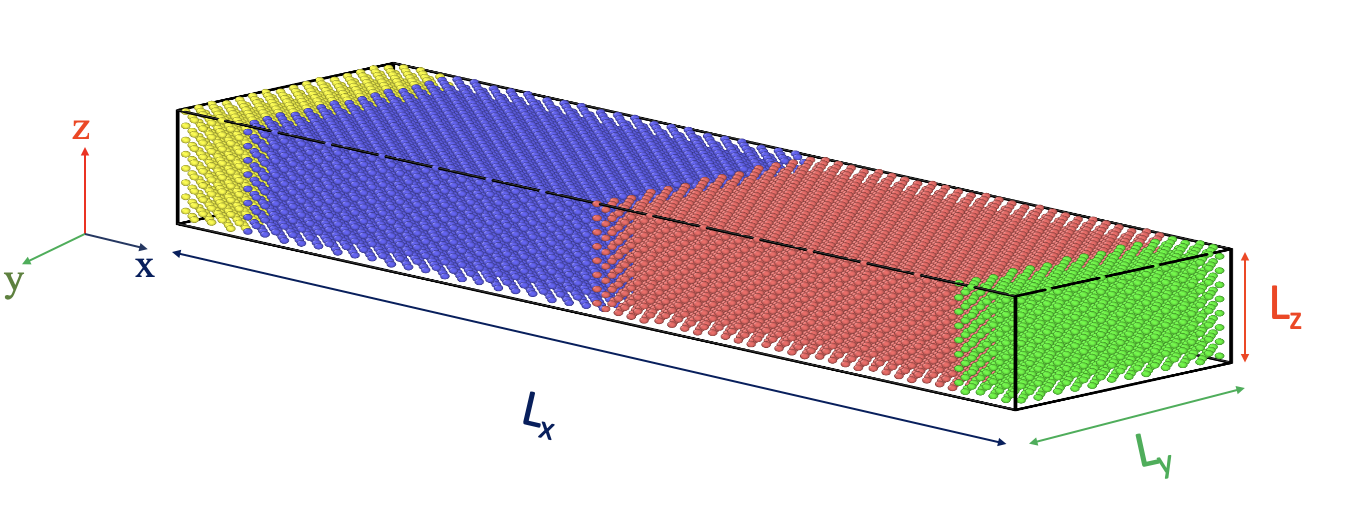}
    \caption{Bicrystal with a STGB used in this study to conduct MD simulations. Blue atoms represent grain 1, red atoms represent grain 2, yellow atoms represent the boundary layer in grain 1 and green atoms represent the boundary later in grain 2. The box is periodic in y and z directions.}
    \label{fig:md_bicrystal}
\end{figure}

In this section, we describe the details of MD simulations used to validate the shear coupling predicted from the energy barriers computed in Step 5. We follow Step 1 in \Cref{sec:stgb_construction} to construct bicrystals with flat ground state GBs. The bicrystals were oriented such that the tilt axis lies along the $z$-axis, the GB plane is situated in the $y-z$ plane, and the GB normal is oriented along the $x$-axis. PBCs were imposed in the $y$ and $z$ directions. The dimension of the simulation cell along the $x$ axis was $L_x = \SI{10}{\nm}$, and the dimensions $L_y$ and $L_z$ along the $y$ and $z$ directions, respectively, were set equal to integer multiples of CSL vectors.  In all simulations $ L_z = \SI{1.5}{\nm}$ (based on the tilt axis), while  $\SI{1}{\nm} \le L_y \le \SI{5}{\nm}$.  We simulated an NPT ensemble with a temperature $\SI{100}{\kelvin}$ set using the Nos\'e--Hoover thermostat and a pressure of 1 atmosphere. The choice of a low temperature for MD simulations was motivated by the fact that our NEB-based calculations are performed at \SI{0}{\kelvin}.  We allowed the GBs to attain the target temperature and equilibrate for $\SI{1}{\ns}$, after which we began loading. \Cref{fig:md_bicrystal} shows a sample bicrystal used in this study to investigate GB coupling factor using MD.

We used two types of forces to drive the system --- (a) shear strain along the $y$-axis on the $yz$-plane  and (b) chemical potential across the two grains. For shear loading, we first fixed a block of atoms at the top and bottom of the bicrystal to serve as boundary layers in the non-periodic direction (the thickness of these blocks was greater than the cutoff radii of the interatomic potentials). We then subjected the system to a constant shear rate of $10^6\si{\per\second}$ by translating the atoms in the top layer at a constant velocity. For chemical potential loading simulations, we applied a synthetic driving force using the \texttt{orient/eco} \cite{Schratt_orienteco} function in LAMMPS. Since we are using low temperatures, the magnitude of the applied chemical potential needs to be high (~$\SI{0.06}{\eV}$/atom).  The coupling factor was calculated by tracking a strip of atoms using a fiducial marker across the $x$-axis while the system was evolved until the GB moved a distance of $\SI{1.5}{\nm}$ along the $x$-axis.

\section{Results}
\label{sec:results}
\begin{figure}[h]
    \centering
    \begin{subfigure}[b]{0.485\textwidth}
        \includegraphics[width=\linewidth]{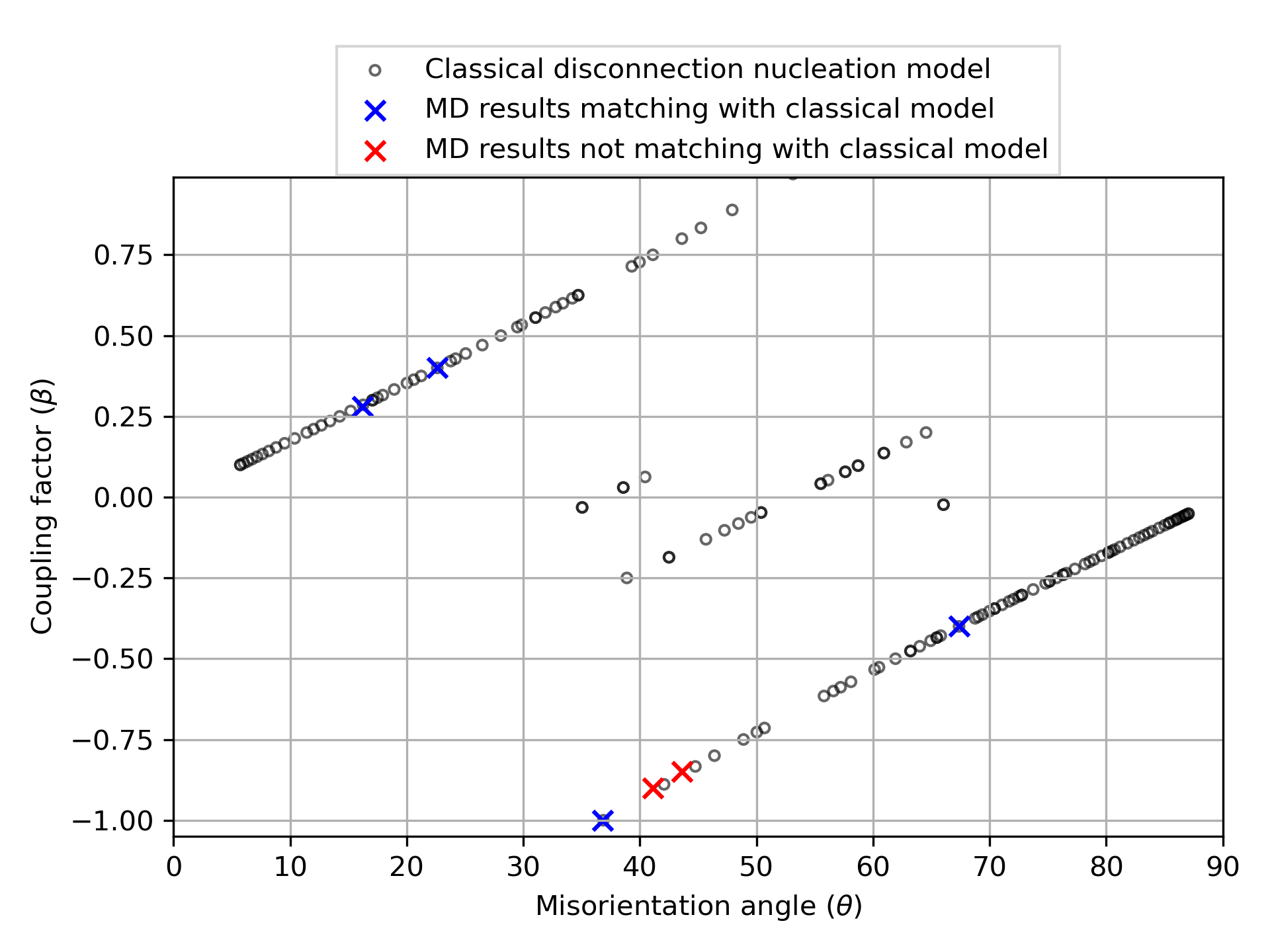}
        \caption{shear force-driven}
        \label{fig:classical_theory_shear} 
    \end{subfigure}
     \begin{subfigure}[b]{0.485\textwidth}
        \includegraphics[width=\linewidth]{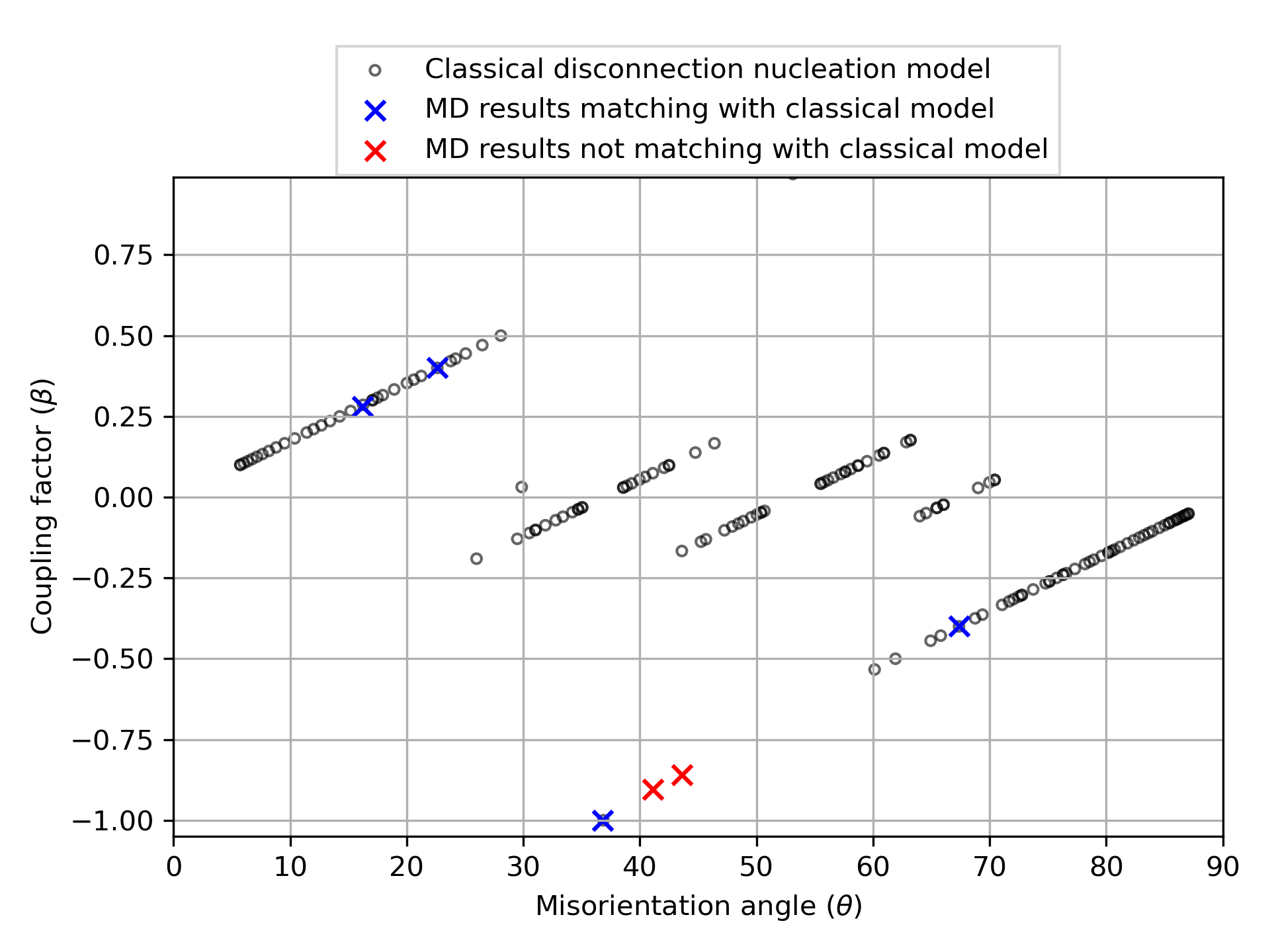}
        \caption{chemical potential-driven}
        \label{fig:classical_theory_chempot} 
    \end{subfigure}
    \caption{A comparison of shear coupling factors of $\hkl[001]$ STGBs predicted by the dislocation-based classical model of \citet{khater2012disconnection} (black circles) and those measured in MD simulations (crosses). The blue crosses represent MD data that align with the predictions of the classical model, while the red crosses signify disagreement. }
    \label{fig:srolovitz}
\end{figure}

We calculated the coupling factor using the \emph{classical model} for high shear stress and applied chemical potential to match the driving forces applied in MD simulations. \Cref{fig:srolovitz} compares the shear coupling factors predicted by the classical model to those observed in MD simulations and shows the agreement is reasonable. However, the presence of outliers at misorientation angles $\SI{43.60}{\degree}$ and $\SI{41.11}{\degree}$ motivates us to employ atomistic methods outlined in \Cref{sec:Methods} to predict the energy barriers and shear coupling factors of the GBs listed in \Cref{tab:GBs} under different loading conditions.

\begin{table}[h!]
    \centering
    \begin{tabular}{|c|c|c|c|c|c|}
    \hline
      \multirow{2}{*}{GB \hkl[001]-tilt}
      & \multirow{2}{*}{$\theta$} 
      & \multicolumn{2}{c|}{$\beta$ - classical} & \multicolumn{2}{c|}{$\beta$ - MD (100 K) }  \\
      \cline{3-6}
      & & shear-driven & potential-driven & shear-driven & potential-driven \\
      \hline
         $\Sigma 13\hkl(5 1 0)$ &  22.62$^\circ$ & 0.40 & 0.40 & 0.40 & 0.40 \\
         $\Sigma 29\hkl(5 2 0)$ &  43.60$^\circ$ & 0.8 & -0.16 & -0.84 & -0.86 \\
         $\Sigma 73\hkl(11 5 0)$ &  41.11$^\circ$ & 0.75 & 0.07 & -0.91 & -0.904 \\
         \hline
    \end{tabular}
    \caption{A list of $\hkl[001]$-STGBs comparing the predictions of the shear coupling factors by the classical disconnection model to those observed in MD simulations.}
    \label{tab:GBs}
\end{table}
 
\subsection{Energy barriers for GB motion}
\begin{figure}[t]
    \centering
    \begin{subfigure}[b]{0.48\textwidth}
    \includegraphics[width=\linewidth]{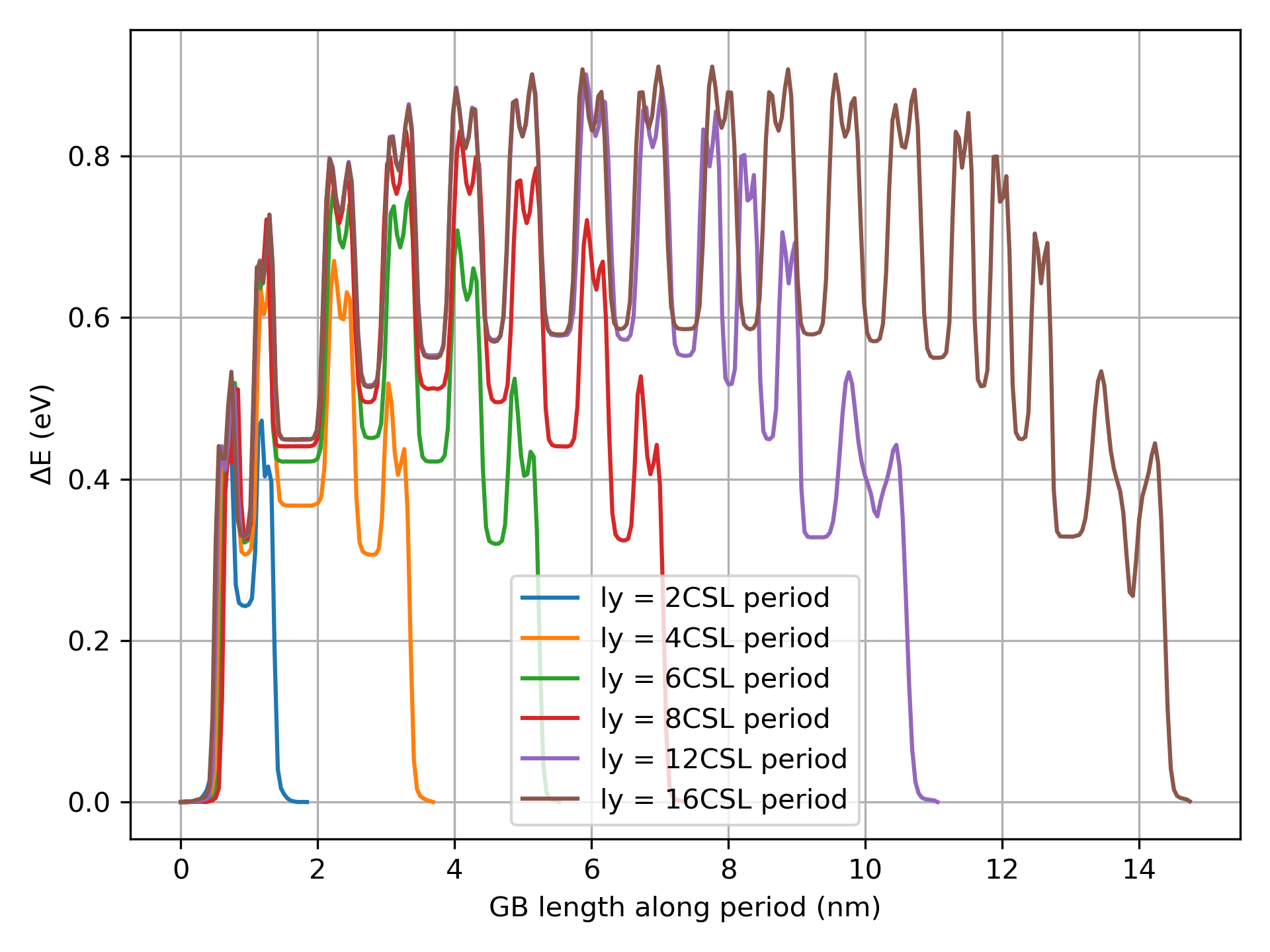}
    \caption{}
        \label{fig:se13} 
    \end{subfigure}
    \begin{subfigure}[b]{0.48\textwidth}
    \includegraphics[width=\linewidth]{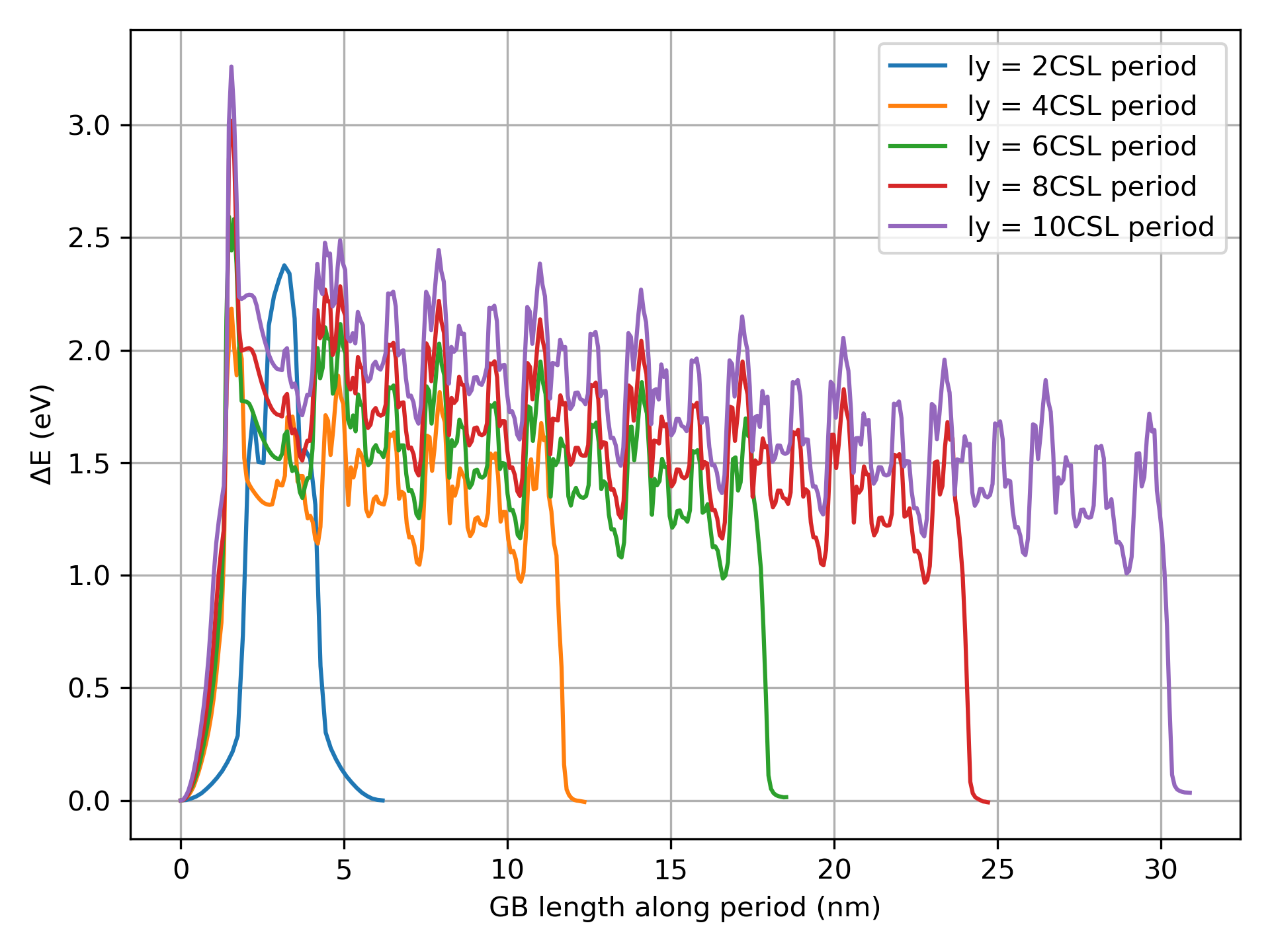}
    \caption{}
        \label{fig:se73} 
    \end{subfigure}
    \caption{Minimum energy path per unit area as a function of the reaction coordinate for multiple box lengths along $y$-axis (a)$\Sigma$13[001] disconnection mode $(0.71,1.77)$ (b)$\Sigma$73[001] disconnection mode $(1.27,1.69)$. As the simulation box size increases, the energy per unit area converges to a value. }
    \label{fig:size_effect}
\end{figure}
To eliminate the influence of image forces in a finite periodic simulation cell on the disconnection nucleation barrier, it is necessary to demonstrate that the simulation box is suitably large.
\Cref{fig:size_effect} shows the 
MEPs for $\Sigma$13[001] disconnection mode $(0.71,1.77)$ and $\Sigma$73[001] disconnection mode $(1.27,1.69)$. From the plots, it is clear that the energy barrier of the MEP converges as the GB length increases. In particular, we note that a GB length of 6 CSL period is sufficient to predict energy barriers within an error of 5\%. Therefore, all GBs hereafter have a length of 6 CSL period. 

\begin{figure}[t]
    \centering
    \begin{subfigure}[b]{0.48\textwidth}
    \includegraphics[width=\linewidth]{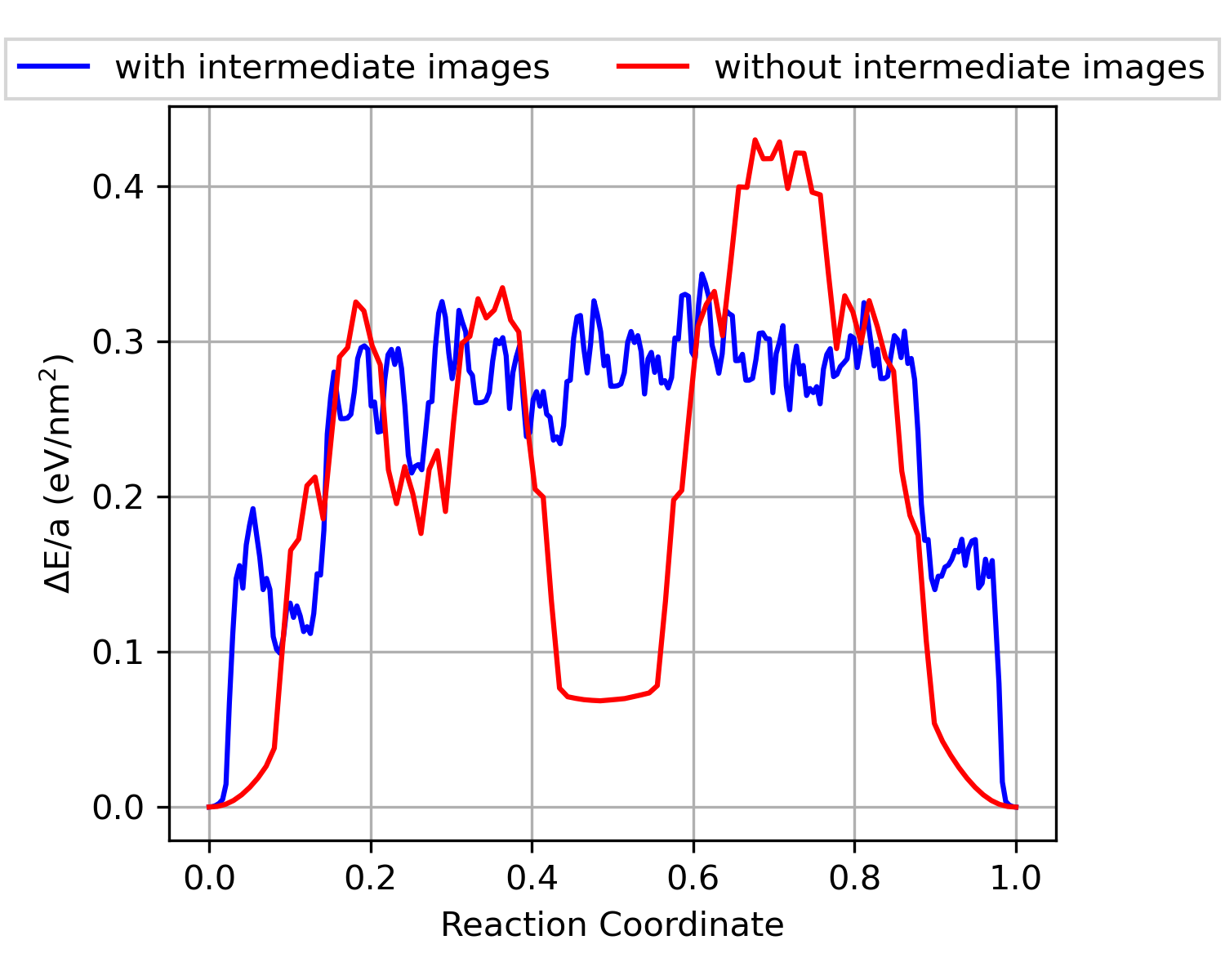}
    \caption{$(b,h) = (0.67,5.71)$}
        \label{fig:se29_mode1_imagecomp} 
    \end{subfigure}
    \begin{subfigure}[b]{0.48\textwidth}
    \includegraphics[width=\linewidth]{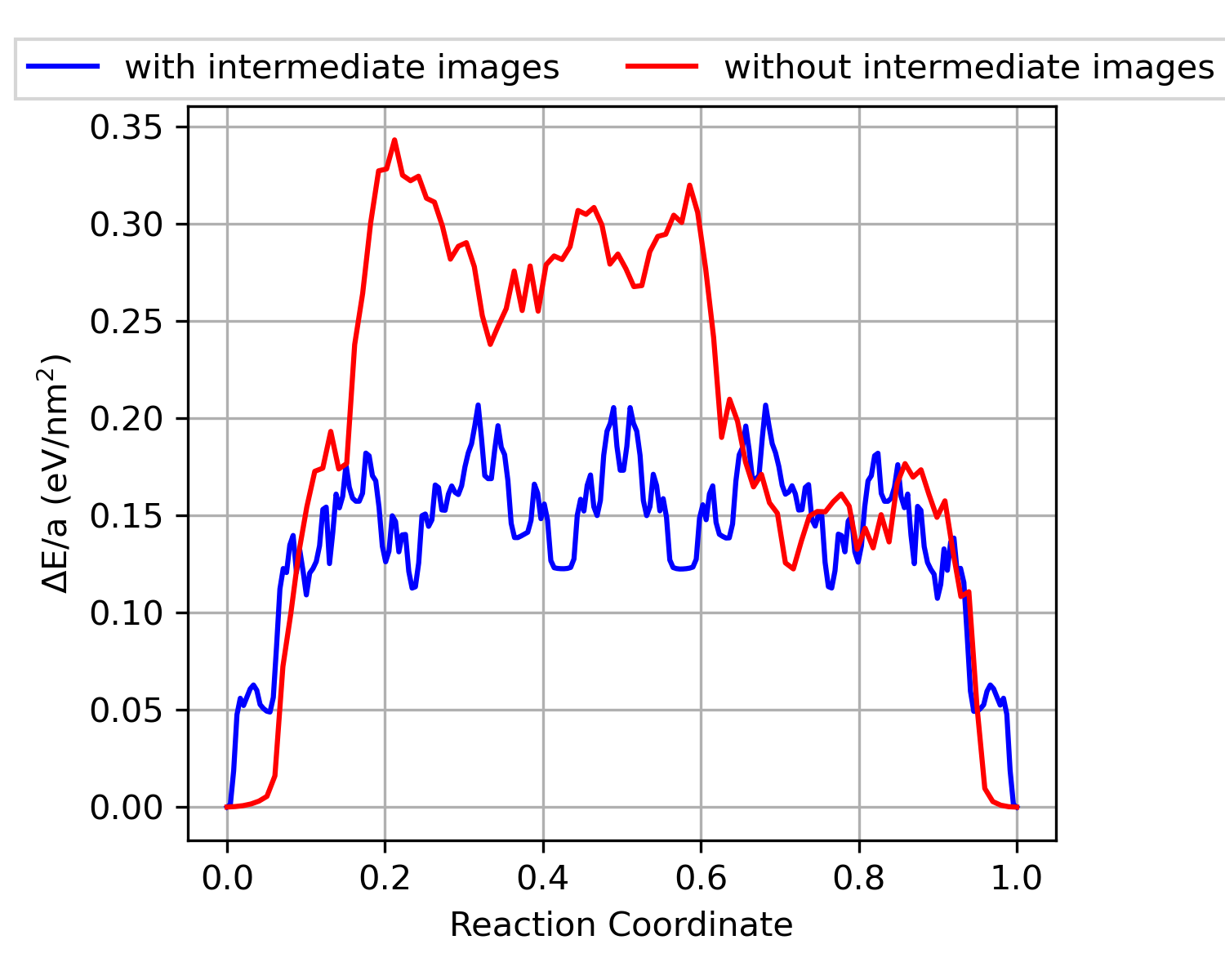}
    \caption{$(b,h) = (0.67,-4.03)$}
        \label{fig:se29_mode2_imagecomp} 
    \end{subfigure}
    \caption{MEP as a function of the reaction coordinate two disconnection modes in $\Sigma 29\hkl[001]\hkl(5 2 0)$. Each plot shows the comparison for MEP calculated by providing intermediate images and not providing intermediate images.}
    \label{fig:imagecomp}
\end{figure}

We also evaluate the effect of providing intermediate images to NEB. To this end, we conduct two calculations: one in which we provide intermediate images to NEB, and the other where we provide only the initial and final images without any intermediates images. \Cref{fig:imagecomp} shows the results of such a calculation for two disconnection modes in $\Sigma 29\hkl[001]\hkl(5 2 0)$. We can clearly observe that the energy barrier calculated is higher for cases where we do not provide the intermediate images compared to those where we do. This can be attributed to the fact that NEB tends to nucleate multiple dislocation dipoles when only the initial and final GB states are provided to it. These dipoles interact with each other, leading to an increase in the energy barrier. However, by providing intermediate images to NEB, we ensure that only one dipole is nucleated and glides along the GB. This results in lower energy barriers. The above observation applies to all but two disconnection modes (see below). 

\begin{table}[t]
    \centering
    \begin{tabular}{|c|c|c|c|c|}
    \hline
      \multirow{2}{*}{GB \hkl[001]-tilt}
      & \multirow{2}{*}{$\theta$} 
      & \multicolumn{3}{c|}{Disconnection mode details}  \\
      \cline{3-5}
      & & $\beta$ & burgers vector,b ($\AA$) & step height, h ($\AA$) \\
      \hline
         \multirow{3}{*}{$\Sigma 13\hkl(5 1 0)$} &  \multirow{3}{*}{22.62$^\circ$} & -0.25 & 0.71 & -2.84 \\
         & & -1.33 & 1.42 & -1.06\\
         & & 0.4 & 0.71 & 1.77\\
         \hline
         \multirow{3}{*}{$\Sigma 29\hkl(5 2 0)$} &  \multirow{3}{*}{43.60$^\circ$} & 0.79 & 1.34 & 1.69  \\
         & & -0.166 & 0.67 & -4.03\\
         & & -0.85 & 2.01 & -2.35\\
         \hline
         \multirow{3}{*}{$\Sigma 73\hkl(11 5 0)$} &  \multirow{3}{*}{41.11$^\circ$} & 0.75 & 1.27 & 1.69  \\
         & & -0.21 & 0.85 & -4.02\\
         & & -0.9 & 2.13 & -2.33 \\
         \hline
    \end{tabular}
    \caption{Disconnection modes studied for the STGBs under consideration.}
    \label{tab:disconnection_mode_details}
\end{table}

\begin{figure}[t]
    \centering
    \begin{subfigure}[b]{0.32\textwidth}
    \includegraphics[width=\linewidth]{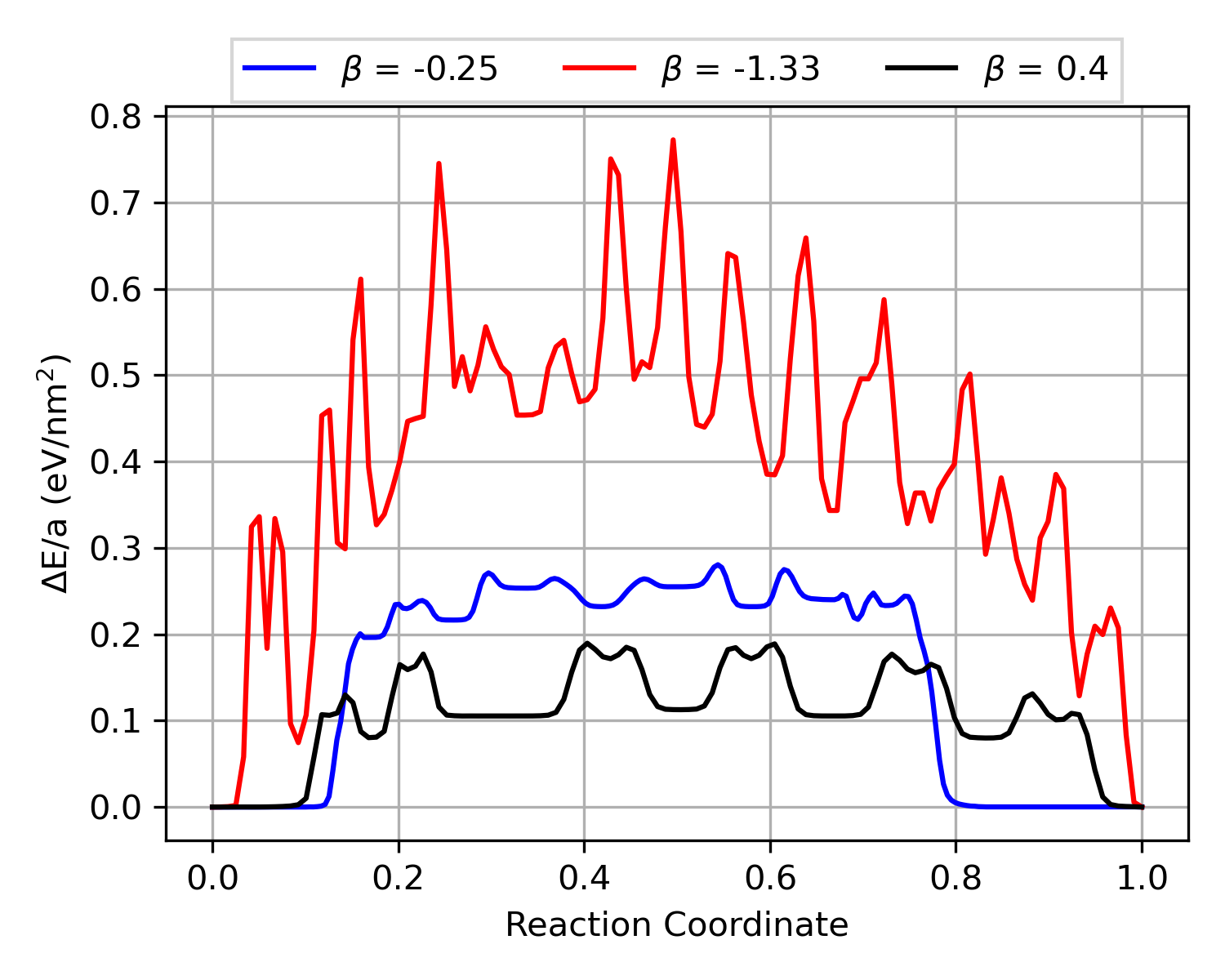}
    \caption{$\Sigma 13\hkl[001]$}
        \label{fig:se13_noload} 
    \end{subfigure}
    \begin{subfigure}[b]{0.32\textwidth}
    \includegraphics[width=\linewidth]{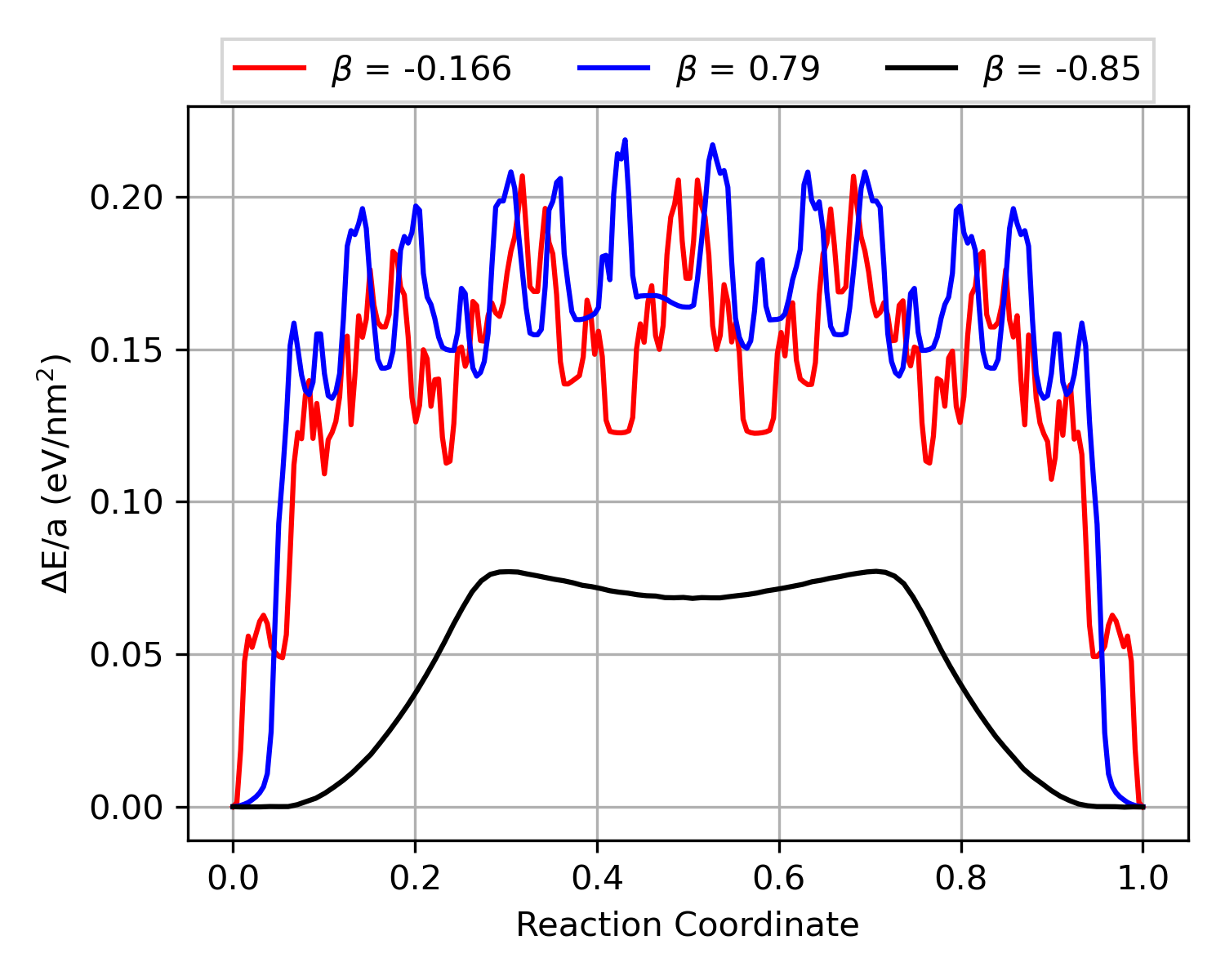}
    \caption{$\Sigma 29\hkl[001]$}
        \label{fig:se29_noload} 
    \end{subfigure}
    \begin{subfigure}[b]{0.32\textwidth}
    \includegraphics[width=\linewidth]{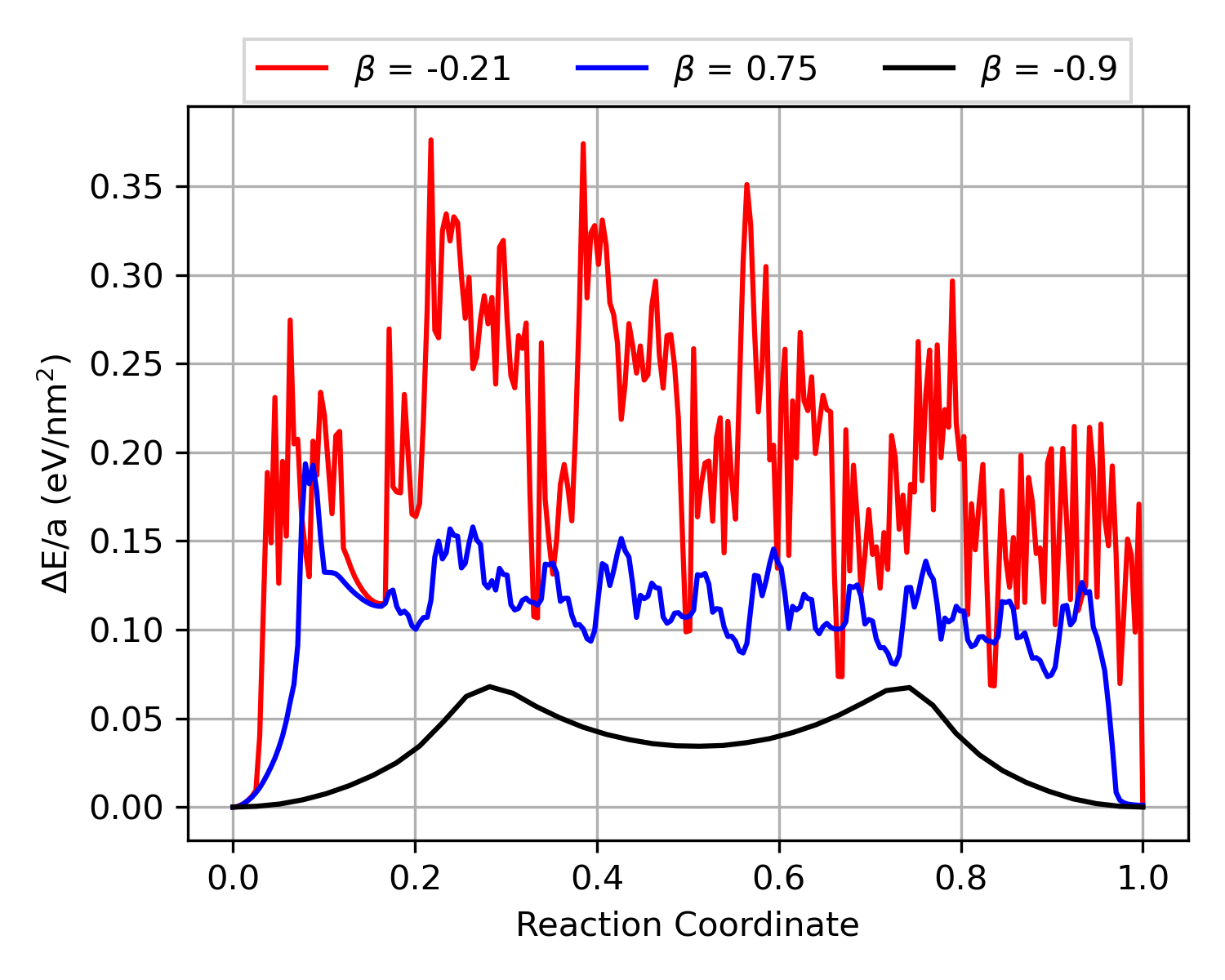}
    \caption{$\Sigma 73\hkl[001]$}
        \label{fig:se73_noload} 
    \end{subfigure}
    \caption{Minimum energy path as a function of the reaction coordinate.}
    \label{fig:noload_barrier}
\end{figure}
Next, we compute MEPs of individual disconnection modes in the three STGBs listed in \Cref{tab:GBs} using Steps 1-5 described in \Cref{sec:stgb_construction}. Although an STGB has countably infinite disconnection modes, we focus our attention on the bottom three modes (see \Cref{tab:disconnection_mode_details}) in terms of the energy barrier predicted by the classical model. 
The plots in \Cref{fig:noload_barrier} show MEPs calculated using NEB, and the resulting atomic trajectories, provided as movies in Supplementary information, show the atomistic mechanism of nucleation and glide of disconnections. The raw data obtained using NEB for each of the disconnection modes and sizes is provided in a Mendelev data repository \cite{supplementary}. The reaction coordinate $RC=0$ represents the initial flat GB, while $RC=1$ represents the flat GB translated by the step height. As we noted earlier in Section \ref{sec:Methods}, constructing images with stable disconnection steps is challenging for disconnection modes with larger burgers vectors. We encountered this challenge in $\Sigma 29[001]$ $(\beta=-0.85)$ and  $\Sigma 73[001]$ $(\beta=-0.9)$ disconnection modes (shown in black in \Cref{fig:se29_noload} and \Cref{fig:se73_noload}, respectively) where the Burgers vector is 3 and 5 times the smallest Burgers vectors, respectively. For these modes, we provide only the initial and final images and no intermediate images. This leads to MEPs with much lower number of local minima than the modes where we provide intermediate images. The intermediate energies represent the barriers for the nucleation and glide of a single disconnection dipole. In comparison, the classical disconnection theory results in a single nucleation barrier as it assumes that nucleation is the rate-limiting step for GB migration, which is a reasonable assumption if the barriers for glide are negligible compared to the nucleation barrier. While the plots corresponding to $\Sigma 13[001]$ $(\beta=-0.25)$, $\Sigma 29[001]$ $(\beta=-0.85)$, and $\Sigma 73[001]$ $(\beta=-0.9)$ in \Cref{fig:noload_barrier} show negligible glide barriers, those corresponding to $\Sigma 13[001]$ $(\beta=-1.33)$, and $\Sigma 29[001]$ $(\beta=-0.166)$ suggest otherwise. The MEPs of the latter modes suggest that disconnection glide is the rate-limiting step. In addition, we note that the disconnection mode in $\Sigma13[001]$ with the lowest energy barrier has the smallest Burgers vector accessible to the GB, while those in $\Sigma29[001]$ and $\Sigma73[001]$ Burger vectors with magnitudes that are 3 and 5 times the smallest Burgers vectors, respectively. This suggests that $\bm b$ or $h$ exclusively do not determine the energy barrier.

\subsection{Shear coupling mode prediction under applied load}
We predict the shear coupling factor under the hypothesis that it is determined by the disconnection mode with the lowest energy barrier. Since the barrier depends on the driving force, our strategy is to calculate the MEPs of disconnection modes in the absence of external loads and then infer the MEPs in the presence of arbitrary driving forces. If $\Delta E$ denotes the MEP calculated in the absence of external loads for a disconnection mode $(\bm b,h)$, then the energy barriers in the presence of shear stress and chemical potential are given by
\begin{subequations}
\begin{align}
    \Delta E_{\rm s} &= \Delta E -\tau a |b| (\delta_h/h) = \Delta E - \tau a \beta \delta_h , \qquad \text{(shear-driven)}  \label{eqn:shear_work}\\
    \Delta E_{\rm c} &= \Delta E -\psi a \delta_h, \qquad \qquad \qquad \text{(chemical potential-driven)} \label{eqn:cp_work}
\end{align}
\label{eqn:work}%
\end{subequations}
where 
$\delta_h$ is the average normal displacement of the GB normal to Gb plane,  $a$ is the area of GB, $\beta$ is the GB coupling factor, $\tau$ is the shear stress applied on the system and $\psi$ is the chemical potential per unit volume applied on the system. The last terms in \cref{eqn:work} represent the work done by the external loads --- chemical potential $\psi$ and shear stress $\tau$.

From \Cref{tab:GBs}, we first note that the classical model and MD predictions of the shear coupling factor ($\beta=0.4$ prediction for shear- and chemical potential-driven cases) agree for $\Sigma 13[001]$ GB. Meanwhile, the coupling factors observed using MD at 100 K for $\Sigma 29[001]$ and $\Sigma 73[001]$ are significantly greater in magnitude than those predicted by the dislocation-based model.

\begin{figure}[t!]
    \centering
     \begin{subfigure}[b]{\textwidth}
        \includegraphics[width=\textwidth]{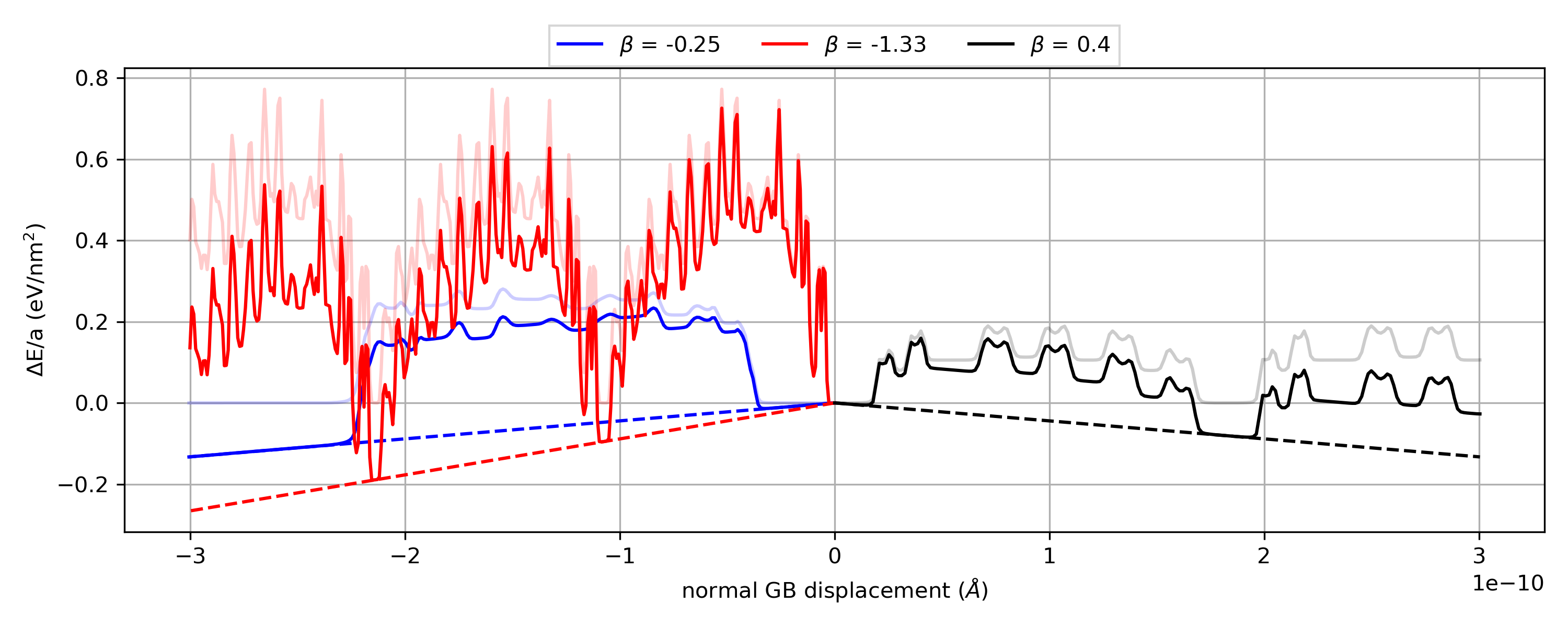}
        \caption{$\Sigma$13[001], 300 MPa shear stress}
        \label{fig:mep13_sh} 
    \end{subfigure}
    \hfill
     \begin{subfigure}[b]{\textwidth}
        \includegraphics[width=\textwidth]{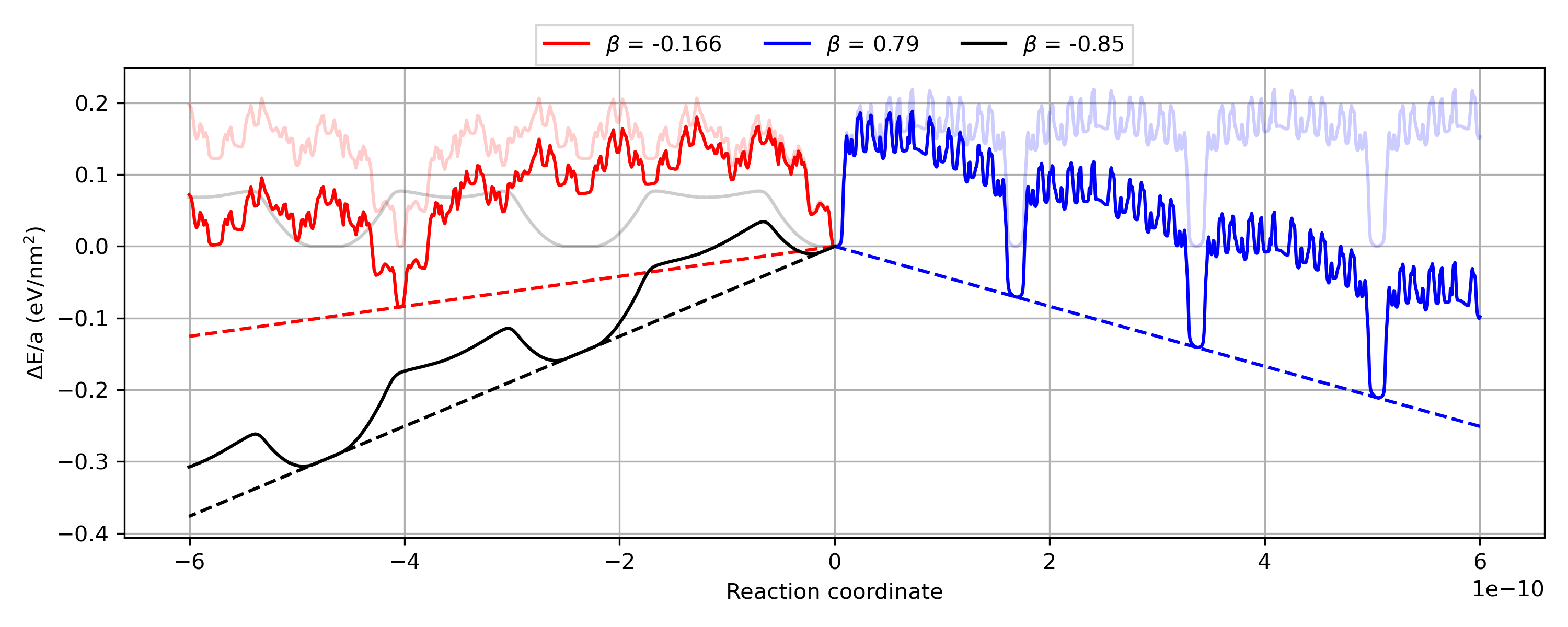}
        \caption{$\Sigma$29[001], 300 MPa shear stress}
        \label{fig:mep29_sh} 
    \end{subfigure}
    \hfill
     \begin{subfigure}[b]{\textwidth}
    \includegraphics[width=\textwidth]{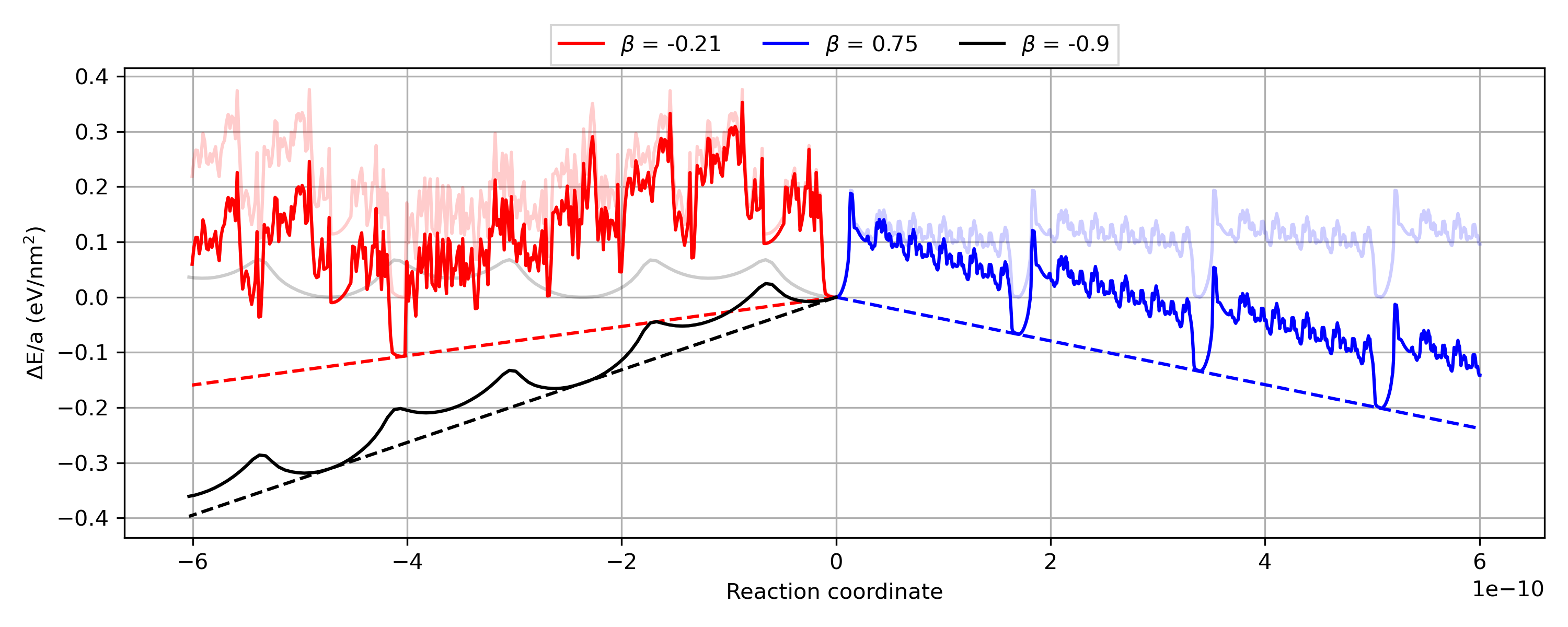}
        \caption{$\Sigma$73[001], 300 MPa shear stress}
        \label{fig:mep73_sh} 
    \end{subfigure}
\caption{MEPs (solid lines) corresponding to shear-driven GB migration, plotted with respect to the normal displacement of the GB. Transparent plots correspond to MEPs under no loading computed using NEB. Dashed lines represent the work done by the shear stress.} 
    \label{fig:mep_shear}
\end{figure}

\begin{figure}[t!]
    \centering
    \begin{subfigure}[b]{0.5\textwidth}
        \includegraphics[width=\textwidth]{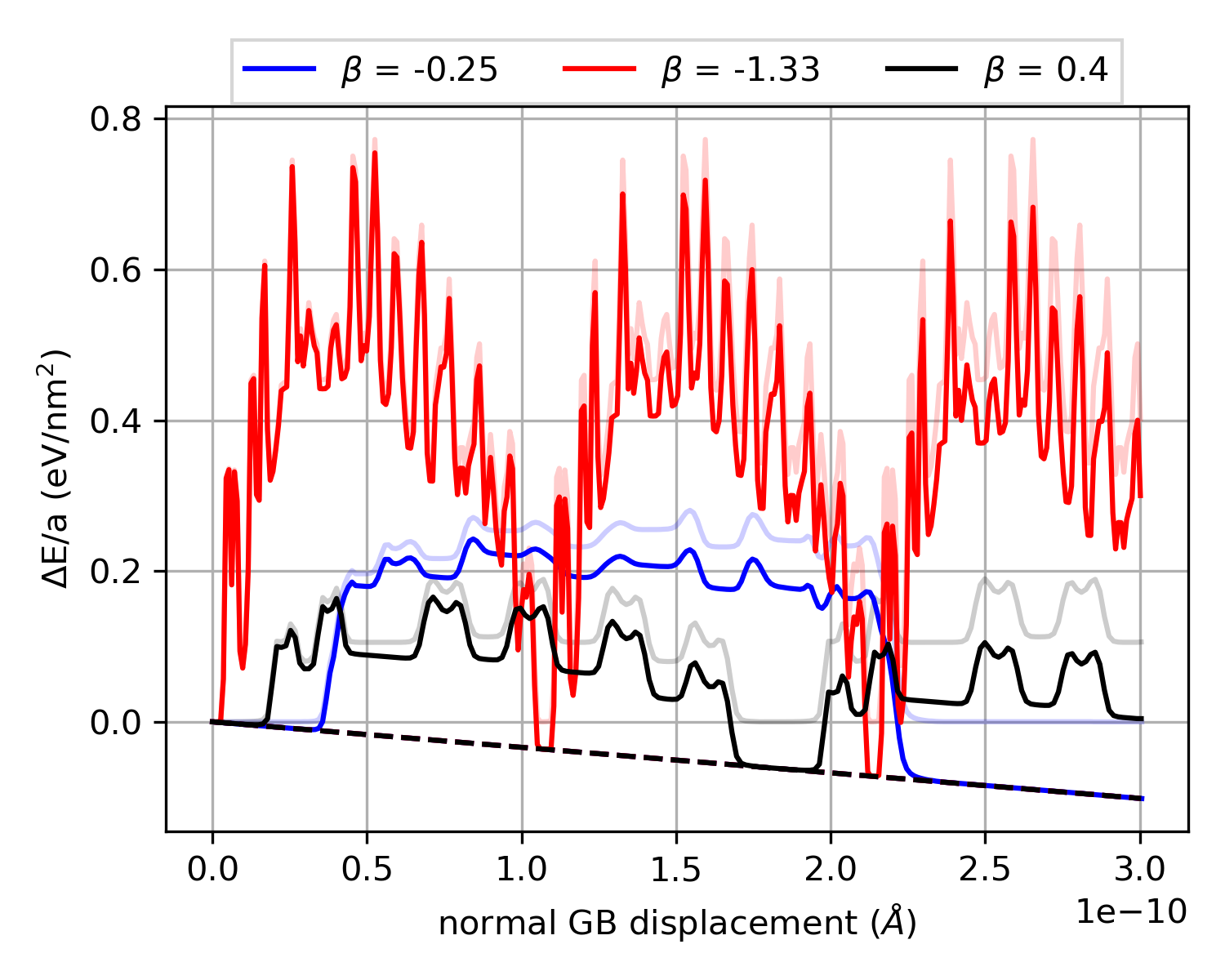}
        \caption{$\Sigma$13[001], 0.04 eV/$\AA^{-3}$ chemical potential}
        \label{fig:mep13_cp} 
    \end{subfigure}
    \hfill
    \begin{subfigure}[b]{0.5\textwidth}
        \includegraphics[width=\textwidth]{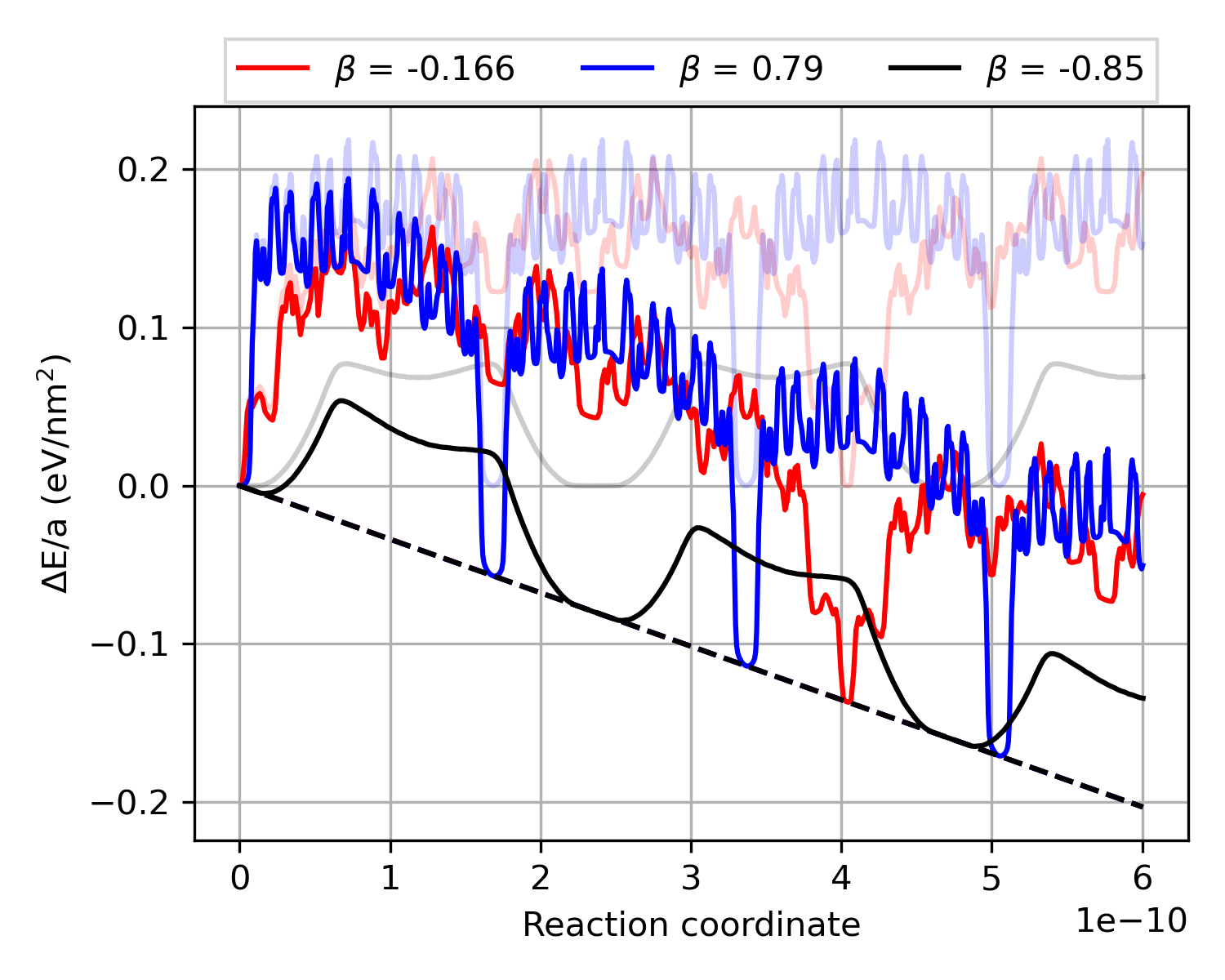}
        \caption{$\Sigma$29[001], 0.04 eV/$\AA^{-3}$ chemical potential}
        \label{fig:mep29_cp} 
    \end{subfigure}
    \hfill
    \begin{subfigure}[b]{0.5\textwidth}
    \includegraphics[width=\textwidth]{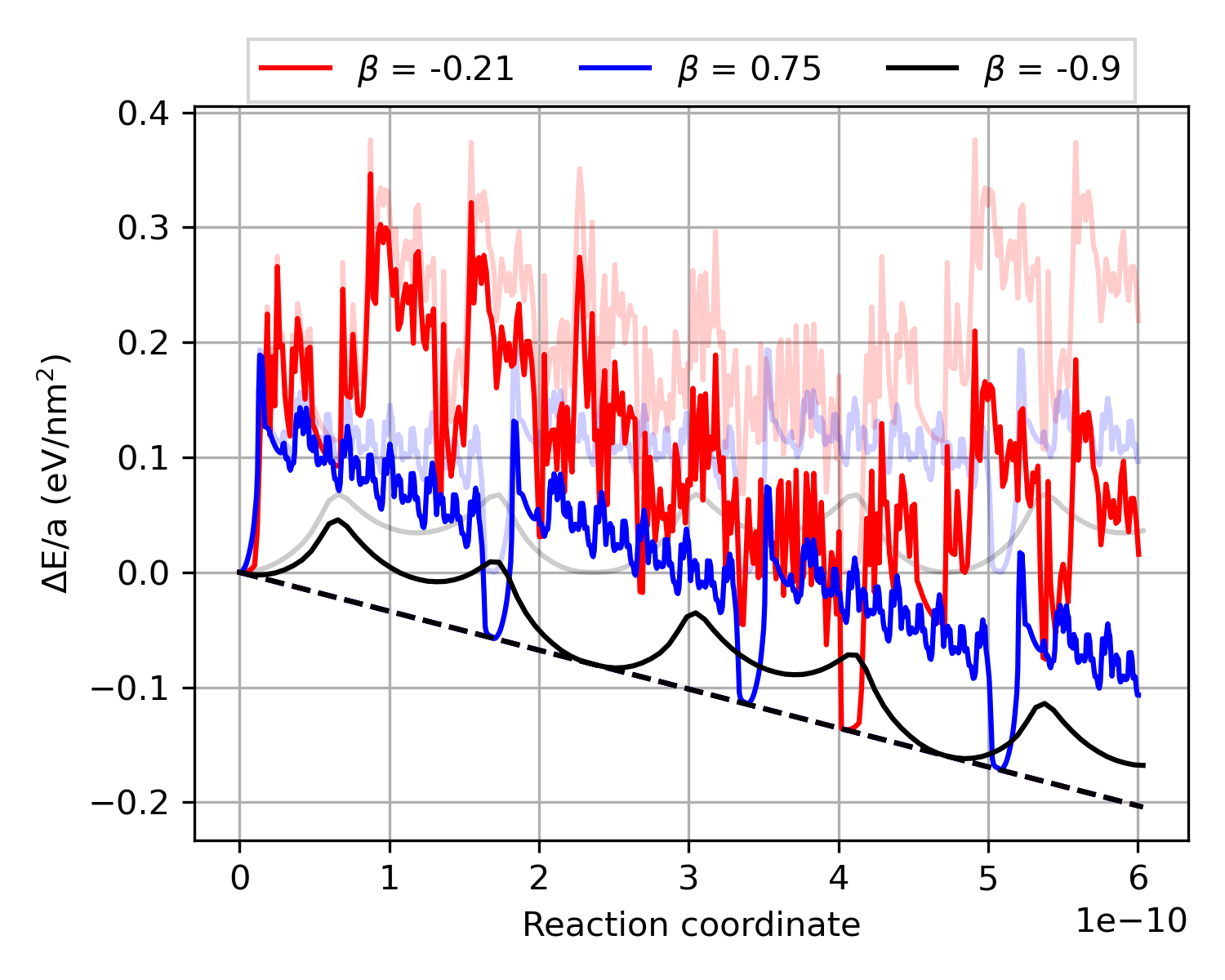}
        \caption{$\Sigma$73[001], 0.04 eV/$\AA^{-3}$ chemical potential}
        \label{fig:mep73_cp}
    \end{subfigure}
\caption{MEPs (solid lines) corresponding to chemical potential-driven GB migration, plotted with respect to the normal displacement of the GB. Transparent plots correspond to MEPs under no loading computed using NEB. Dashed lines represent the work done by the synthetic driving force.} 
    \label{fig:mep_cp}
\end{figure}

\Cref{fig:mep_shear} and \Cref{fig:mep_cp} display the MEPs plotted against the normal displacement of the GB for shear stress ($\tau=\SI{300}{\mega\pascal}$) and chemical potential ($\psi=\SI{0.04}{\eV}\AA^{-3}$) driven cases, respectively. To facilitate a comparison between the plots, we chose a constant normal displacement for each of the GBs --- $3 \AA$ for $\Sigma13[001]$ and $6 \AA$ for $\Sigma29[001]$ and $\Sigma73[001]$). Each figure consists of three sets of plots --- 1) the transparent plots show the MEP $\Delta E$ calculated using the NEB method for a free boundary; 2) the dotted lines represent the work done by the external force, which is linear in $\delta_h$ and with a slope equal to $\tau a \beta$ in the shear-driven case and $\psi a$ in the chemical potential-driven case. (see \cref{eqn:shear_work} and \cref{eqn:cp_work}) the solid plots, obtained as a sum of the corresponding previous two plots, represent the MEP of disconnection modes driven by an external force.

In the shear-driven case (\Cref{fig:mep_shear}), the MEP of a disconnection mode with a positive shear coupling factor lies along the positive $x$-axis, indicating that the GB moves upward when subjected to a positive shear load. Similarly, those modes with a negative $\beta$ have their MEPs along the negative $x$-axis. On the other hand, in the chemical-driven case (\Cref{fig:mep_cp}), MEPs of all modes are along the positive $x$-axis since the GB always moves upward in the direction of the applied force.

\Cref{fig:mep_shear} and \Cref{fig:mep_cp} illustrate that in each of the GBs considered, the mode depicted in black has the lowest energy barrier when subjected to both shear stress and chemical potential. These modes align with the coupling factor observed in molecular dynamics (MD) simulations at low temperatures. For $\Sigma 29[001]$ and $\Sigma 73[001]$, energy barrier calculations based on the NEB method provide a more reliable pathway to predict the coupling factor than classical dislocation theory. Notably, the proposed method accurately predicts the coupling factor of STGBs in low-temperature regimes.  In our investigation of higher temperature regimes for $\Sigma 29[001]$ and $\Sigma 73[001]$, we find that the coupling factor changes significantly as the temperature increases, with the magnitude of the coupling factor decreasing at elevated temperatures. At temperatures exceeding 300 K, the magnitude of the coupling factor decreases with increase in temperature, and around 600 K the GBs show a coupling factor close to that of the mode represented in red in \Cref{fig:mep_shear} and \Cref{fig:mep_cp} under chemical potential application. Since diffusion continues to be limited at 600 K, we expect that the shear coupling continues to be disconnections mediated. However, the energy that determines the observed disconnection mode will be the free energy barrier instead of the internal energy barrier calculated in this work. 

\section{Conclusion}
\label{sec:Conclusions}
In this paper, we develop an atomistic framework to calculate the energy barriers associated with the nucleation and glide of individual disconnection modes in grain boundaries (GBs) subjected to shear and/or chemical potential. Based on the hypothesis that the disconnection mode with the lowest energy barrier determines the shear coupling factor, the framework was used to predict the shear coupling factors in $\hkl[001]$ STGBs, including those for which the predictions of the classical disconnection model do not agree with the MD simulations. The two key components of our framework are --- a) the construction of individual disconnection modes using Smith normal form bicrystallography and b) the analysis of their energetics using the NEB method with atomic trajectories adherent to the \emph{min-shuffle} mapping. Consequently, a unique advantage of our framework is that by mapping the minimum energy paths (MEPs) of individual disconnection modes in a free (no driving forces) GB, we were able to predict the shear coupling factor for \emph{any} driving force.

Our predictions for coupling factors depend on the magnitude of the nucleation energy barriers obtained from the calculated MEPs. The developed method is accurate for low-temperature regimes (T$ \le 300 $K), where the predictions made using this method align with the MD results. However, at higher temperatures, GBs exhibit coupling factors of lower magnitude than those predicted by this method. We will reserve the use of these low-temperature barriers to predict coupling factors at higher temperatures for future work.  The framework also yields the atomistic mechanism and migration barriers included in the MEPs, which opens numerous avenues to explore non-Arrhenius GB migration and stress/chemical potential-driven GB phase transitions and to inform continuum disconnection models, such as the ones developed by \citet{han2022disconnection} and \citet{joshi2022finite}.

\section*{Supplementary information}
\label{sec:Supplementary}
The ancillary information contains 9 movies showing mechanisms of GB migration corresponding to the modes tabulated in \Cref{tab:disconnection_mode_details} and their MEPs shown in \Cref{fig:noload_barrier}.

\section*{CRediT authorship contribution statement}

\textbf{Himanshu Joshi}: Data curation, Formal analysis, Investigation, Methodology, Software, Validation, Visualization, Writing – original draft. \textbf{Ian Chesser}: Investigation, Supervision, Writing – review \& editing. \textbf{Brandon Runnels}: Investigation, Project administration, Resources, Writing – review \& editing. \textbf{Nikhil Chandra Admal}: Conceptualization, Formal analysis, Funding acquisition, Investigation, Methodology, Project administration, Supervision, Writing – review \& editing.

\section*{Acknowledgments}
NCA and HJ would like to acknowledge support from the National Science Foundation, grant $\#$ MOMS-2239734. BR acknowledges support from National Science Foundation, grant $\#$ MOMS-2341922. IC acknowledges support from the LANL Postdoc Program via the Metropolis Fellowship.

\bibliographystyle{elsarticle-num-names}
\bibliography{ref.bib}

\end{document}